REVIEW

# Towards Time Sensitive Networking on Smart Cities: Techniques, Challenges, and Solutions

**Rui Eduardo Lopes**[1,2] | **Duarte Raposo**[2] | **Susana Sargento**[1,2]

[1]Universidade de Aveiro, 3810-193 Aveiro, Portugal

[2]Instituto de Telecomunicações, 3810-193 Aveiro, Portugal

**Correspondence**
Rui Eduardo Lopes
Email: ruieduardo.fa.lopes@ua.pt

Funding Information
This research was supported by the Fundação para a Ciência e Tecnologia within grant FCT.2021.06223.BD, with DOI: 10.54499/2021.06223.BD, and by the PRR – Plano de Recuperação e Resiliência and the NextGenerationEU funds at University of Aveiro, through the scope of the Agenda for Business Innovation "NEXUS: Pacto de Inovação – Transição Verde e Digital para Transportes, Logística e Mobilidade" (Project nº53 with the application C645112083-00000059).

## Abstract

Smart cities transform urban landscapes with interconnected nodes and sensors. The search for seamless communication in time-critical scenarios has become evident during this evolution. With the escalating complexity of urban environments, envisioning a future with a blend of autonomous and conventional systems, each demanding distinct quality-of-service considerations, services in smart cities vary in criticality levels and necessitate differentiated traffic handling, prioritizing critical flows without compromising the network's reliability or failing on hard real-time requirements. To tackle these challenges, in this article, we discuss a time-sensitive networking approach, which presents multi-faceted challenges, notably interoperability among diverse technologies and standards at the scale of a smart city network. TSN emerges as a promising toolkit, encompassing synchronization, latency management, redundancy, and configuration functionalities crucial for addressing smart city challenges. Moreover, the article scrutinizes how TSN, predominantly utilized in domains like automotive and industry, can be tailored to suit the intricate needs of smart cities, emphasizing the necessity for adaptability and scalability in network design. This survey consolidates current research on TSN, outlining its potential in fortifying critical machine-to-machine communications within smart cities while highlighting future challenges, potential solutions, and a roadmap for integrating TSN effectively into the fabric of urban connectivity.

## 1 | INTRODUCTION

The number of smart cities is increasing worldwide, with the number of people who live in cities rising dramatically in recent years, mostly from 2017 [121]. Several heterogeneous computational nodes and sensors are displaced to support the deployment of systems, covering the largest possible area of such urban locations.

Once static entities, cities are dynamic ecosystems where technological advancements shape their core. Central to this evolution is smart cities, where an intricate web of interconnected elements coexist within the urban fabric, striving to elevate the quality of life for their inhabitants.

With the proliferation of smart cities and populations within these social and technological contexts, the probability of unwanted events occurring, such as accidents, complex vehicle traffic conditions, and higher pollution levels, increases and develops the need to communicate information describing them. Moreover, considering evolved scenarios in which autonomous elements (such as vehicles) coexist with conventional elements, networks capable of distinguishing time-critical tasks and traffic from everyday events become an urgent requirement. Some cities worldwide already have services capable of transmitting an example of such content to people, such as emergency-oriented messages or alerts [54]. Nevertheless, and as we are detailing a smart city feature, ideally, one should not only want to communicate data directly to people but, more importantly, to other computational nodes for them to react and execute their procedures to these events without overwhelming citizens with impractical alerting scenarios. This article then attempts to study how critical flows in machine-to-machine (M2M) communications can be transmitted whilst ensuring requirements for its hard real-time characteristics.

### 1.1 | Motivation

For a long time, the scientific community has designed and implemented ways of controlling and allowing differentiated traffic to be transmitted according to a priority [8]. In a smart city environment, many producers and subscribers of services require different quality features, and some, as they strongly differ in their criticality, require features not only to be considered by the network elements but primarily ensured never to fail.

Considering a smart city concept, we should first understand what a smart city will mean some years from now [107]. Our vision contemplates a mixed environment of autonomous and conventional systems (such as vehicles) whose sensed data is acquired from the mobile nodes and the city's communicating sensors. These nodes, providing citizens with multiple functions differing in their quality of service, can establish feedback or communication with people or other machines in a M2M connection, where some time and network requirements are to be placed.

With our experience developing the smart city infrastructure and deploying its services in Aveiro, Portugal [192], this vision seems realistic, embodying our motivation. To better illustrate these concepts, figure 1 depicts our motivation via some use-case scenarios in which different flow examples require latency, reliability, synchronization, and configuration features. Considering the use case scenarios in figure 1 as an example, four different critical flows and city elements reactions are stated as follows.

1) **Ambulance on a mission with reserved lane passage**
Imagine an ambulance on a mission, navigating its way to the hospital [131]. The city efficiently manages traffic to ensure the ambulance's unobstructed passage. This involves restricting regular vehicles from crossing intersections regulated by traffic lights [61, 146], while autonomous vehicles collaborate on their maneuvers to maintain clear lanes for the ambulance [171]. In such a scenario, precise time synchronization of mobile nodes is vital. This ensures that commands are transmitted and received promptly for execution across all participant nodes. This synchronization encompasses mobile nodes within vehicles and roadside units or sensors, such as traffic lights. Furthermore, the traffic flows need to adhere to specific requirements. This enables the elements involved to respond swiftly to emergencies, even during regular operational functions. For example, a traffic light might be scheduled to receive an emergency flow at a predetermined time slot while performing its usual functions. During normal circumstances, the traffic light behaves as expected without an emergency. However, it seamlessly transitions into an emergency context upon receiving an emergency flow. To facilitate this seamless coordination, vehicular messages transmitted through ITS-G5, C-V2X, or similar technologies come into play. Smart cities have already integrated such messages into their interconnection logic [149, 192], connecting both roadside units (RSUs) and on-board units (OBUs) within the urban environment, as well as some of their users, especially vulnerable road users.

2) **Telehealth and remote features from a hospital to an ambulance**
Consider again an ambulance maintaining real-time communication with a hospital, sharing critical patient data and instructions during a remote session with a medical assistance or intervention entity [194]. In this particular scenario, precise synchronization among all participants is paramount. Every command and piece of data must align seamlessly with medical assistance actions. Any deviation in receiving a patient's data could potentially endanger the patient's life. Additionally, considering the various types of data being transmitted—from essential patient data to video streams or a crucial flow of commands—these data streams must adhere to specific latency requirements to prevent disruptions or irregularities. Moreover, to bolster the system's reliability in this scenario, redundant connections or data pathways should be activated to ensure secure delivery. It is equally important to prepare the mobile node's connectivity points along the ambulance's route to the hospital, further fortifying the reliability of the entire system.

3) **Infrastructure safety provisioning after a gas leakage detection**
Imagine a technician vehicle on its way to a location where a gas leak has occurred. The city swiftly designates this area as a high-priority zone, directing emergency services to computational nodes nearest to the incident location [148]. In this setup, the smart city capitalizes on a preceding scenario where traffic lights and autonomous vehicles proactively assist the technician vehicle in reaching the critical location. However, what is genuinely paramount is the intelligent allocation of essential services in this scenario. The smart city does not only leverage the aforementioned cooperative efforts but also optimally assigns public alert systems and other utilities to this location. The aim is to mitigate the risk of fire outbreaks escalating the danger on-site and threatening nearby residents. To achieve this, the nodes in the affected area must be configurable regarding their traffic schedules. Additionally, the communication between nodes should prioritize reliability to ensure the seamless delivery of critical content, whether commands or video streams. This enables remote management of the situation by other machines or units, further enhancing the overall response efficiency.

4) **City asynchronous event triggering safety authorities and services**
A sudden gunshot reverberates through a street, immediately triggering an alert in the vicinity and causing street lights to intensify, enhancing nighttime visibility [233]. Once again, in this scenario, it is paramount that the network of interconnected nodes operates in perfect synchronization, capable of promptly receiving high-priority requests activated by emergencies. This necessitates the nodes to maintain a reliable schedule that consistently accommodates the reception of emergency messages, potentially altering the standard behavior of these devices.

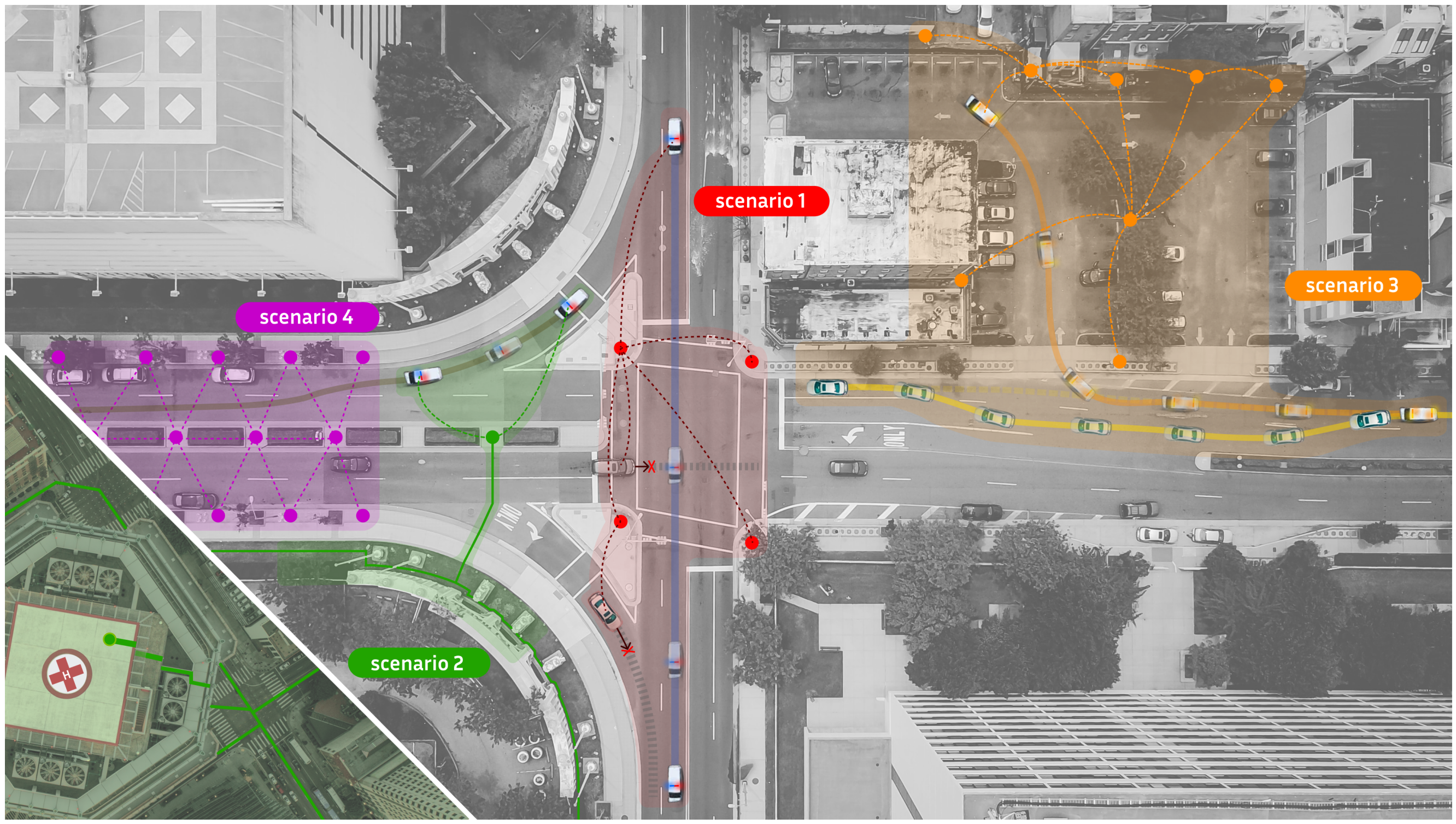

**FIGURE 1** Our motivation supported by four use case scenarios: 1) ambulance on a mission with reserved lane passage; 2) tele-health and remote features from a hospital to an ambulance; 3) infrastructure safety provisioning after a gas leakage detection; 4) city asynchronous event triggering safety authorities and services.

To address such use case scenarios, the industry-recommended toolkit of standards and mechanisms, known as time-sensitive networks (TSNs), was considered [206, 56]. Time-sensitive networks consider various requirements like synchronization, latency, reliability, and configuration features that paved their way onto a toolbox of standards and mechanisms. These TSN functionalities enable us to handle the creation and management of traffic time schedules, time synchronization, redundancy, and other reliability features, ensuring system availability, and allowing mixed-criticality traffic. Additionally, they provide the means to configure new plans for these functions without compromising the current system's service guarantees [87].

However, despite recognizing TSN as a potential solution for the challenges posed by smart cities, implementing it on a large scale, such as a smart city network, remains a challenge [157]. This is in contrast to its common application in scenarios like intra-vehicle networking, as closed-circuit M2M industrial network communications. The smart city use case for TSN can be seen as a large-scale integration of smaller and more commonplace use cases for TSN found in domains like automotive [26], industry [114], avionics [152], and others.

Interoperability is crucial to integrating small TSN systems and other network approaches and systems within a smart city. This allows the diverse range of technologies involved to signal themselves and communicate with each other. For instance, smart cities in Europe can benefit from ETSI's ITS-G5 standards [63] and in the US from the equivalent SAE J2735 standard [195], as mentioned in some of the use case scenarios earlier, to enhance urban mobility features. This enables vehicles to communicate functions between them and with the city infrastructure, such as road-side units and traffic lights.

Nevertheless, similar to the ongoing effort with ETSI's ITS-G5 within a smart city, other analogous technologies like WiFi, LoRa, 5G-and-beyond cellular networks are still in the process of integration or standardization as a profile. Moreover, despite profile considerations in both 5G and other wireless technologies, there is a lack of traffic shaping techniques in their backhauls: although existing some quality-of-service (QoS) considerations, these lack make the application of service-based rules difficult. Incorporating these technologies would empower the smart city network to create alternative and redundant backbone paths for the communication of emergency messages. Further details on this topic are covered in this document, mainly focusing on the state-of-the-art.

An extensive array of variables is critical regarding emergency events, time being the primary concern in all types of occurrences.

For instance, with the possibility of cities being large hubs for autonomous vehicles to perform their functions while independent from one another, the city infrastructure could be an answer to disseminating emergency messages concerning the safety of people inside and surrounding the vehicles in timely strict actions. To achieve this goal, deploying networks in smart city contexts, where emergency events are considered, must be done so that prioritized traffic can always have availability guarantees [120].

Regarding emergency events, several countries have already implemented geo-localized emergency alert systems [173]. Such systems can dispatch an emergency-coded message to all recipients within the range of a selected set of mobile network cells proximate to the emergency site. However, these systems depend on the recipients to be devices able to capture such alerts, which do not ensure to entirely cover the population comprised by an event.

## 1.2 | Contributions and Outline

This article is a survey of research on TSN brought into the context of smart cities for M2M communications. As TSN is a recently standardized set of tools (in 2018 [111]), this work also tries to establish a base knowledge of some of the available standards to answer the following questions further:

1. Briefly, what is TSN and what it tries to address?
2. What is the current focus of the scientific community for the usage on TSN, main open challenges, and possible solutions?
3. What are the current and future challenges in smart city computational systems, and how can time-sensitive networking help address them?
4. How can a TSN approach be designed and configured to accommodate a smart city's set of needs?

This survey aims to be a contribution both to the smart city community as well as to the TSN community. It details a new network profile of a smart city for this communications approach that distinguishes from others in its openness regarding both its implementation techniques and entitled frameworks, as well as its network topology, which is far from standard closed-circuit networks found in industrial or automotive infrastructures, as common use cases for TSN.

Following this section, a brief introduction to the methodology used for collecting the state-of-the-art in this paper is presented in section II. Consequently, the state-of-the-art is presented through four main categories of time synchronization, reliability, latency, and configuration, from section III to section VI, respectively. Throughout these sections, the scientific community work is presented in a structured manner, by standard, followed by a discussion where the collected references are strategically presented under the scope of a smart city scenario from our motivation in figure 1. In section VII, a discussion on conclusions and our perspective of possible future work is described.

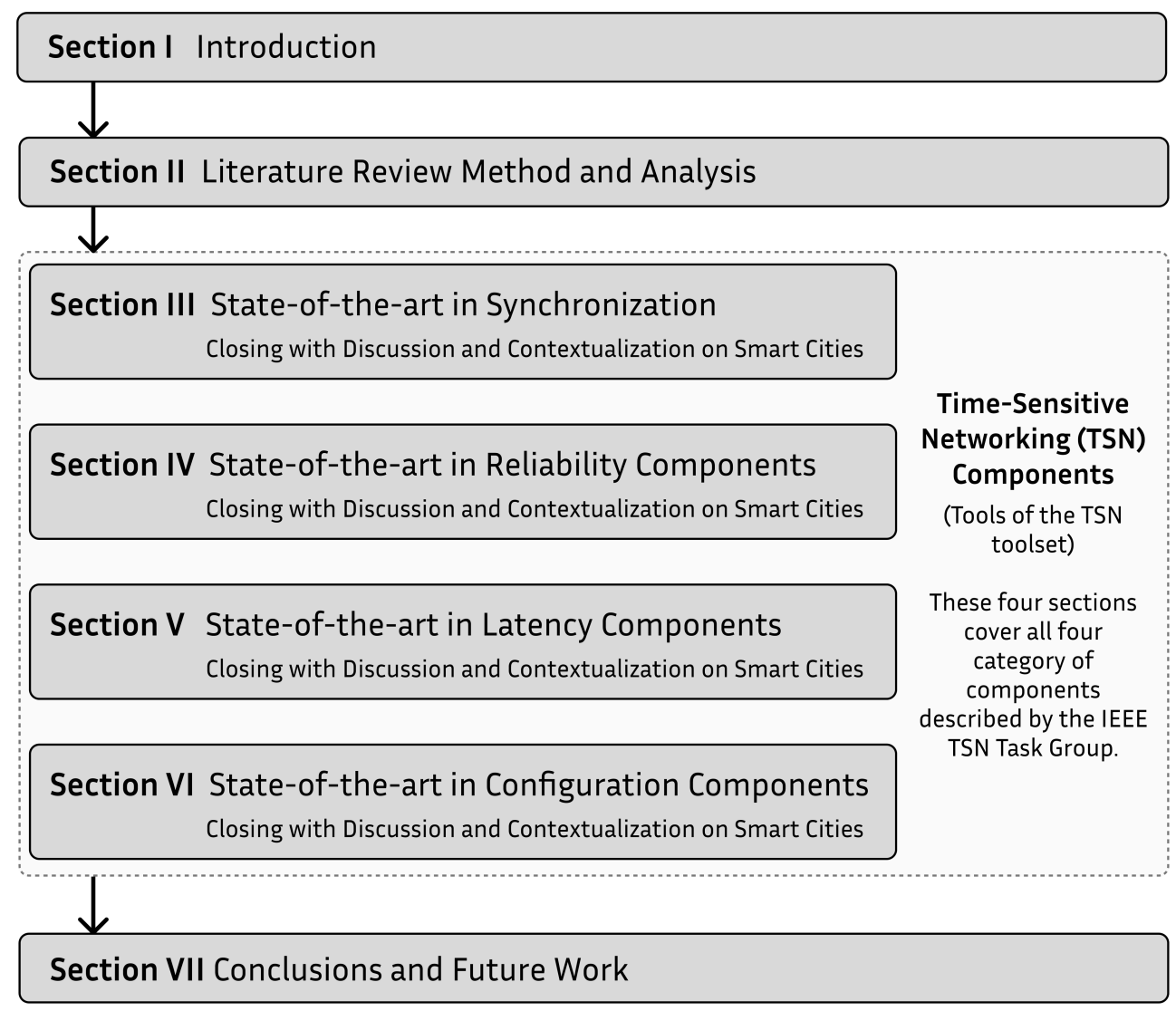

**FIGURE 2** Document outline at a glance. The categorization in four classes of TSN components is adapted from a one-page summary[†], as published by the IEEE TSN Taskgroup.

## 2 | LITERATURE REVIEW METHOD AND ANALYSIS

Concerning the TSN core functions stated before, several works have already been developed by the scientific community for almost 13 years now (since 2012, when the TSN task group was formed in Institute of Electrical and Electronics Engineers (IEEE)). Although 13 years have passed since the task group creation, several works were found to be made on audio/video bridging (AVB) or time-triggered Ethernet (TTE), which were both predecessors in logic to TSN. Although some works are still relevant in this survey, others were innocently incorrect because they were based on discussions and terms from drafts presented pre-2018, as stated by some authors [250, 206].

This section first presents an analysis of the current state-of-the-art relative to both time-sensitive networks and the use case context of smart cities. This study was performed by searching keywords of "TSN", "Time-Sensitive Network", "Time-Sensitive Networking" and other composed keywords with the names of the standards/drafts of TSN, such as "TSN Time-Aware Shaper" or "TSN TAS". Moreover, articles were also searched by joining the context of "smart cities" with it. In total, 246 research publications since 2011 were found and selected in the academic platforms of Google

[†] https://www.ieee802.org/1/files/public/docs2024/admin-tsn-summary-1124-v01.pdf

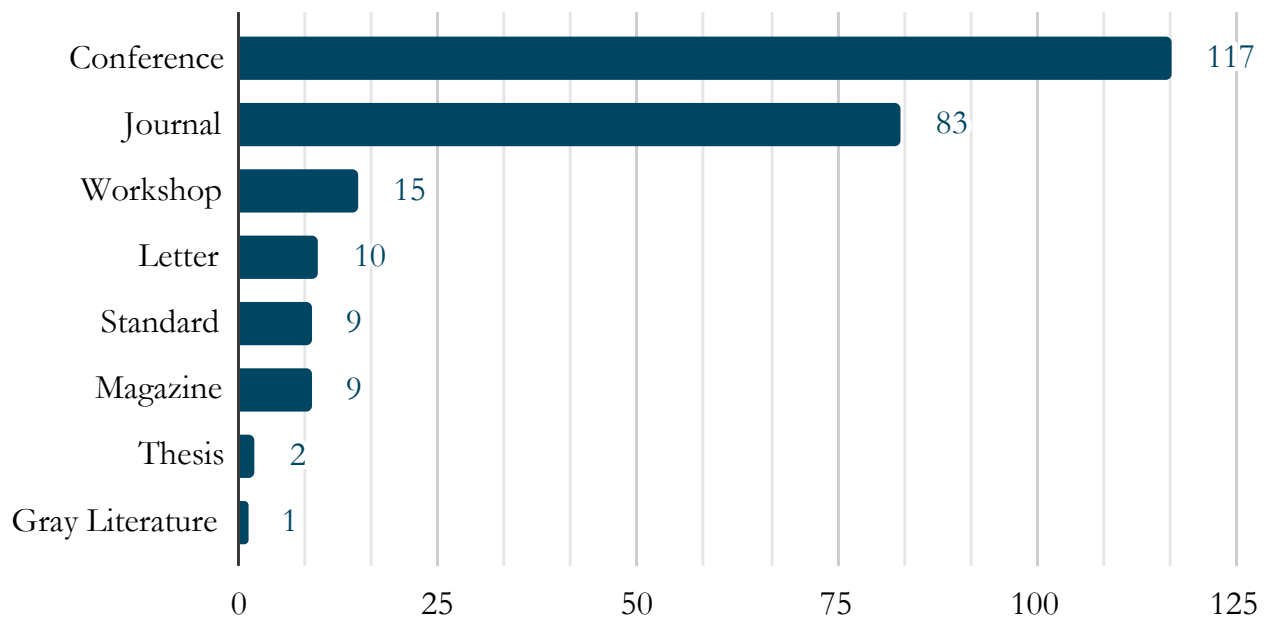

**FIGURE 3** Covered articles in this state-of-the-art per type.

**FIGURE 5** Evolution of use case applications in the selected surveys through time. The selected 2020 surveys are not included here since they do not cover TSN usages or studies.

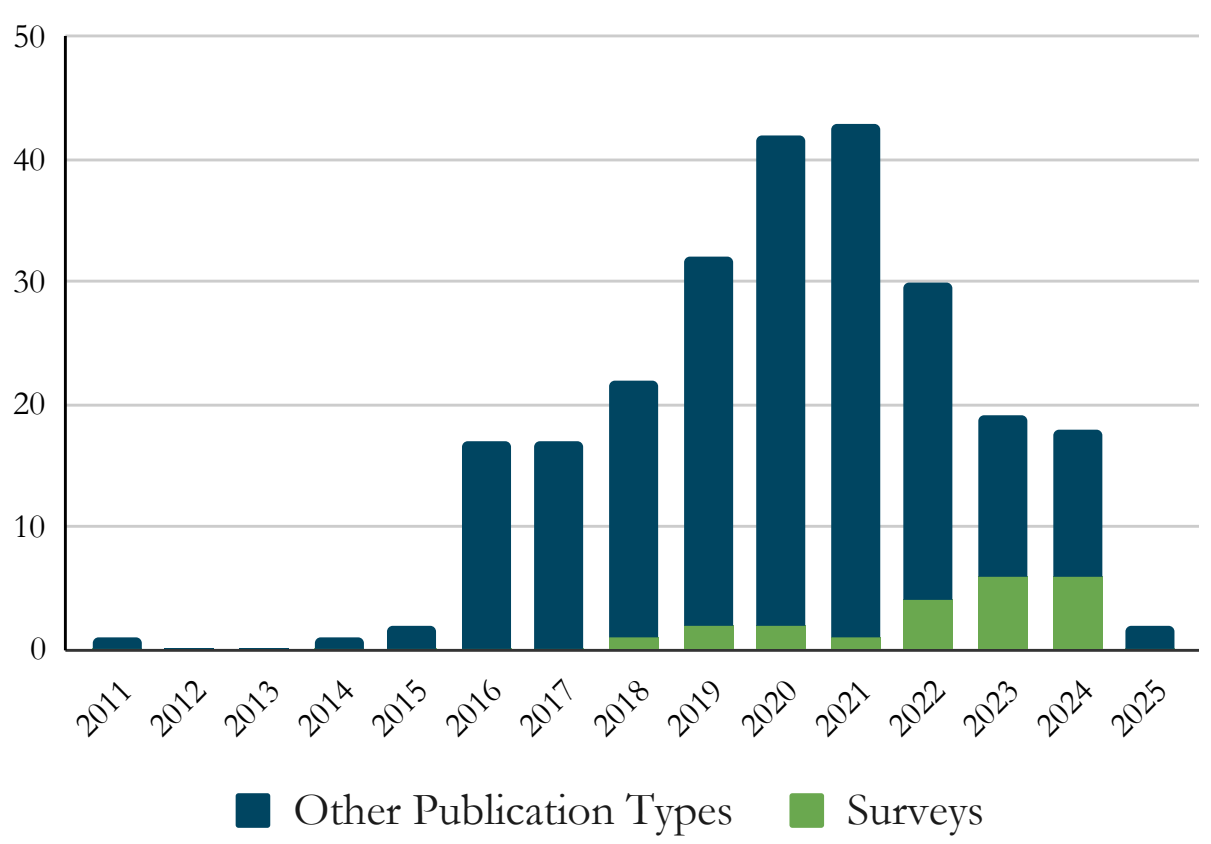

**FIGURE 4** Covered articles in this state-of-the-art. The surveys from the selected articles are highlighted in green.

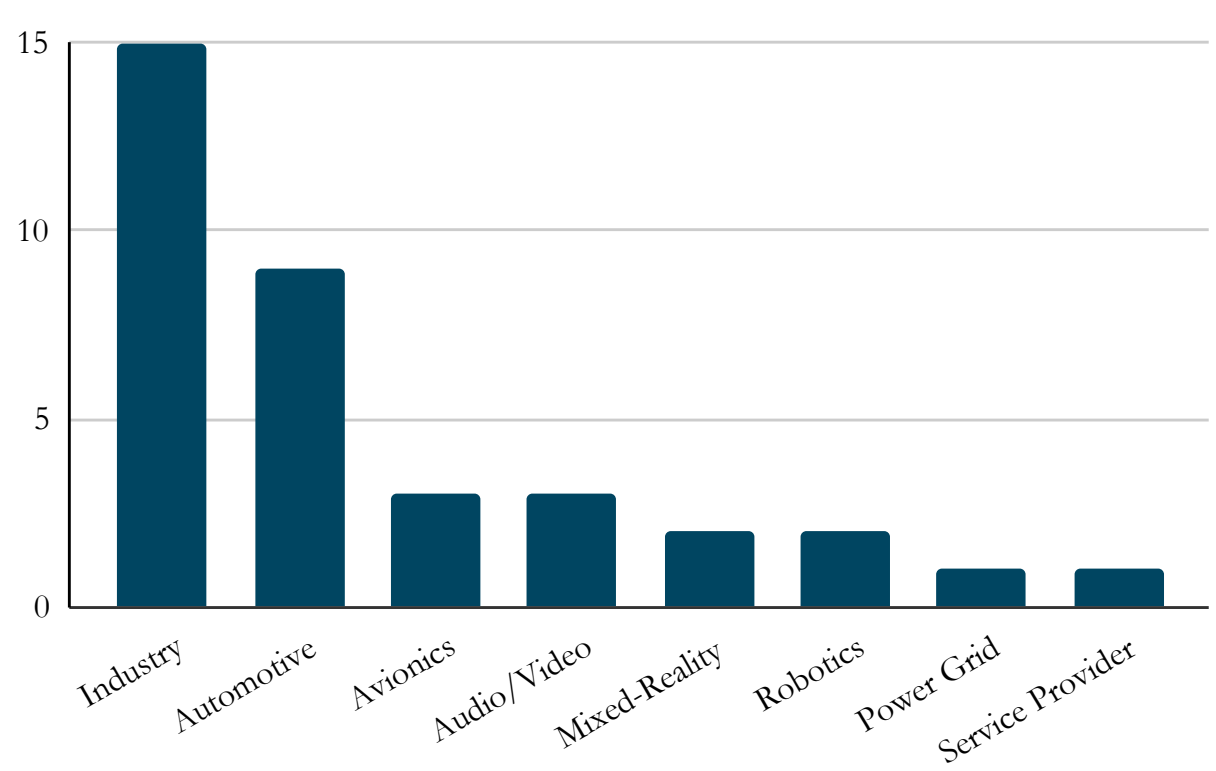

**FIGURE 6** Use case applications of TSN within the selected surveys

Scholar[†], Scopus[‡], and IEEE Xplore[§], with the types described as in figure 3.

From figure 3, both conferences and journals have the majority of targeted publication types, and there are already some workshops that quickly are capable of showing TSN features, mostly under industrial use case scenarios, as authors primarily targeted IEEE's International Workshop on Factory Communication Systems and International Conference on Communication Workshops events. In a more timely manner, and as this article intends to survey the current state-of-the-art regarding TSN and the context of smart cities, in figure 4 a distribution of the collected articles through the years (since 2011) is depicted, in which the surveys are highlighted to showcase the year they had collected information.

From the analysis of figure 4, the number of published articles started to rise more significantly after 2018, following the standardization of TSN in the IEEE 802.1Q-2018. In statistics, in the current year of 2025, two articles were already found, but no conclusion from them in this respect can be made since we are in January at the time of writing.

Regarding the collected surveys, through figure 4, it is easy to affirm that a slight rise in results started sometime after the peak in 2021, whose tendency shows it to be positive. This could be justified by an increase in the search for new alternative application scenarios where time-sensitive networks could be a partially- or fully-functional solution. To support this statement, we have made figure 5 to show the evolution of the use case applications in the selected surveys from 2018 to 2024.

Figure 5 depicts the diversity of use case applications that have evolved through time, proving that TSN is currently being used to tackle other problems rather than the tailored-made industrial, automotive, and audio/video profiles. In absolute value, figure 6 shows the number of times the use case applications have shown in the selected surveys.

† https://scholar.google.com
‡ https://scopus.com
§ https://ieeexplore.ieee.org

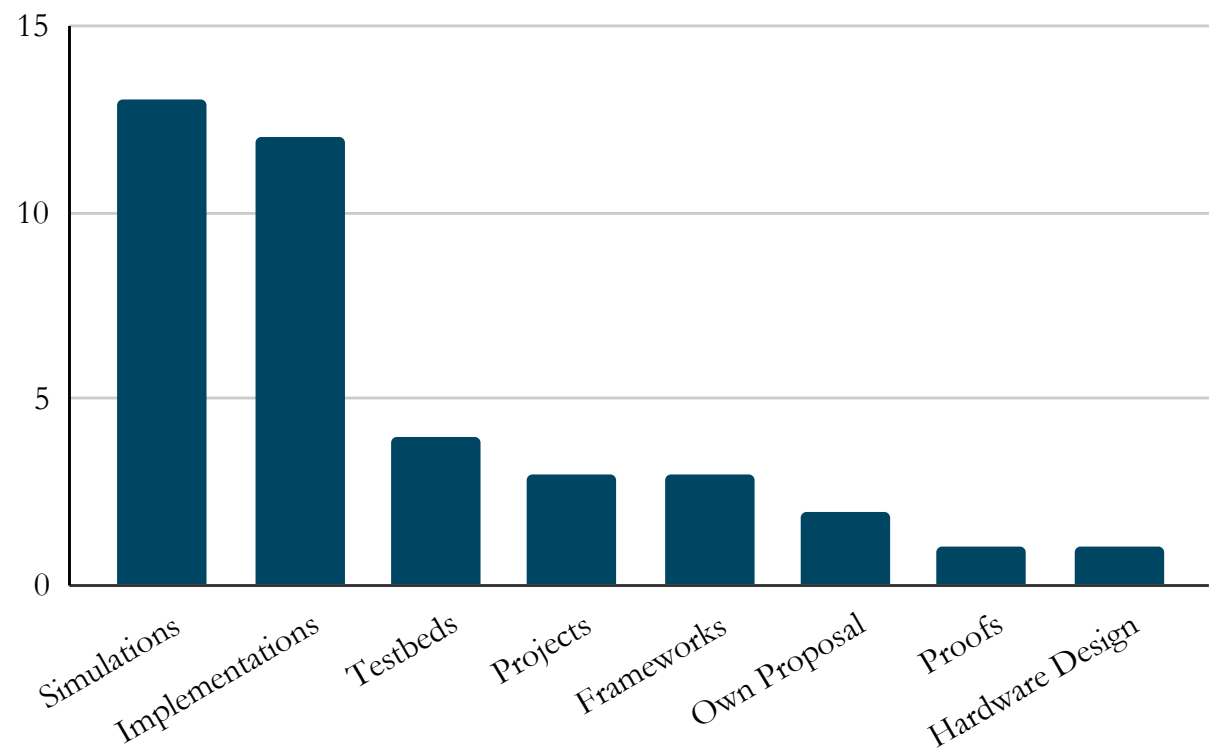

**FIGURE 7** Reported and collected outputs in the selected surveys

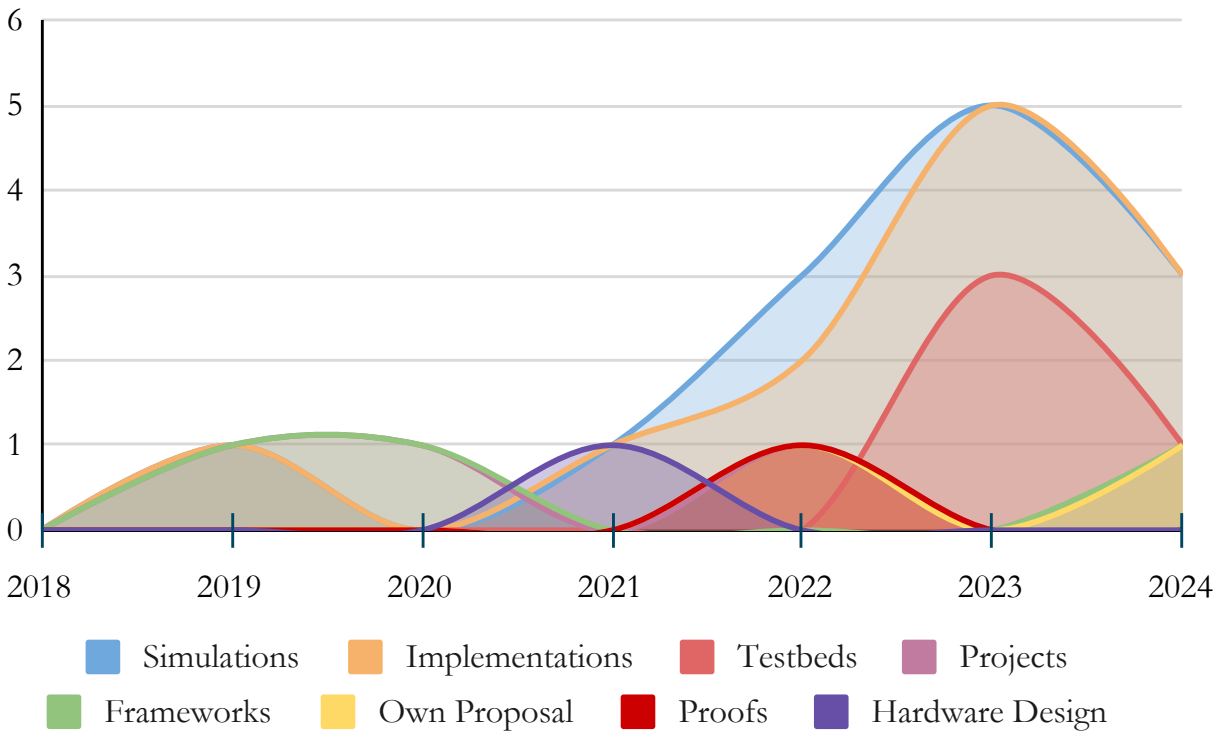

**FIGURE 8** Evolution of the reported and collected outputs in the selected surveys through time.

Whilst the most common application of TSN is industry, which naturally follows the original reason for this toolbox of standards, automotive is the second most common application. Following, and on fewer occasions, avionics appear, following mixed-reality for both augmented and virtual reality applications, where wireless network integration is also required. Between the latter two applications, audio and video are also mentioned, which also traces back to the main original branch of TSN that was AVB. In less number, the scenarios of robotics (as cyber-physical systems), power grid, and service providers are also considered.

Also relevant while characterizing the collected and selected surveys are the types of outputs whose articles redirect to, whether they are original research proposals, proofs, projects where TSN can be found in use as a solution to a given challenge, among other artifacts. Figure 7 shows the absolute number of articles where the given output types can be found.

Similarly to the analysis of found use case applications, the reported and collected output types can also be distributed from 2018 to 2024. As depicted in figure 8, it is visible that the number of simulations and implementations finally started to be concurrent from 2023, as well as the testbeds, that had accompanied the same tendency. The publication of frameworks also increased in 2024, and the result is comprehensible with the number of available simulation tools and implementations made available. To summarize these survey findings, the list of articles is shown in table 1, followed by a set of items that characterize them.

Table 1 not only includes survey references on TSN-relatable studies but also has five references that exclusively cover smart city issues that report open challenges in regards to both sensing and network contexts [60]: sensing-wise, heterogeneous networks are a commonly mentioned challenge to handle; and network-wise, the interconnection between different network access protocols and technologies are a usual issue, followed by the need of fog or edge computing to help complex and numerous city-wide network functions to run in parallel. The incorporation of network infrastructures within the context of smart cities led these environments to follow similar approaches as with Internet of things (IoT) architectures [84, 115], and following the characteristics described as smart mobility and environment by Kirimtat et al. [121].

The creation of urban-sized network infrastructure environments or simulated testbeds is also the primary concern of studies such as Amodei et al. [5]. Nonetheless, a direct relation between both contexts of TSN and smart cities as an application is still pioneering. Despite not being a survey article, the first external reference relating both concepts was found by Yigit et al. [239], whose integration is done under the umbrella of a digital twin in a 6G-enabled network infrastructure of a smart city. Being a hypothesis that we had already defended in another publication [140], the interconnection between TSN and the smart city could offer the latter a strong set of tools able to meticulously configure and manage the diverse array of quality of service levels and resources according to the available network functions. For this reason, table 1 ends with two rows that point out a first journal article whose contribution was to introduce TSN concepts and its usual application use case scenarios, contextualizing them within the network infrastructure and functions of a smart city; and this article, as a comprehensive survey gathering the current state-of-the-art of TSN in regards to what could be used and considered for deployment in a smart city network infrastructure.

The current state-of-the-art will be presented in the sections below following the main composition of TSN by its four main components of synchronization, reliability, latency, and configuration. This study is then summarized in the last section to evaluate the evolution of this subject study and its relevance in the scientific community.

**TABLE 1** List of article surveys from 2018 to 2024, with detail level on tutorials on core TSN concepts, application usages, and integration with other network technologies.

| Survey | Year | Core TSN | Applications | Integration with other technologies |
|---|---|---|---|---|
| Zhao et al. [250] | 2018 | Light tutorial | Industry and Avionics | *none* |
| Nasrallah et al. [163] | 2019 | Full tutorial | Industry and Automotive | 5G and Wi-Fi |
| Seol et al. [206] | 2021 | Full tutorial | Industry, Automotive, Power Grid, and Robotics | Wi-Fi, 5G, and ZigBee |
| Deng et al. [56] | 2022 | Full tutorial | Automotive and Audio/Video | *none* |
| Fedullo et al. [71] | 2022 | Full tutorial | Industry | 5G and Wi-Fi |
| Gavriluţ et al. [81] | 2022 | Light tutorial | *none* | *none* |
| Xin et al. [234] | 2022 | Light tutorial | Industry | 5G |
| Chahed et al. [42] | 2023 | Full tutorial | Industry, Automotive, Avionics, Audio/Video, and Service Provider | *none* |
| Grigorjew et al. [90] | 2023 | Full tutorial | Industry | *none* |
| Maletić et al. [147] | 2023 | Full tutorial | Industry, Automotive, and Audio/Video | *none* |
| Peng et al. [177] | 2023 | Full tutorial | Automotive | CAN and LIN |
| Satka et al. [200] | 2023 | *none* | Industry | 5G |
| Shi et al. [210] | 2023 | Light tutorial | Industry and Automotive | *none* |
| Stüber et al. [219] | 2023 | Light tutorial | *none* | *none* |
| Walrand et al. [231] | 2023 | Full tutorial | Industry, Automotive, Avionics, and Mixed-Reality | *none* |
| Sasiain et al. [199] | 2024 | Full tutorial | Industry | 5G and 6G |
| Xue et al. [235] | 2024 | TAS exclusively | Industry | *none* |
| Zanbouri et al. [243] | 2024 | Full tutorial | Industry, Automotive, Robotics, and Mixed-Reality | 5G, 6G, mmWave, and 802.11be |
| Zhang et al. [248] | 2024 | Full tutorial | Industry | 5G and Wi-Fi |
| Du et al. [60] | 2019 | | *Smart City related survey article, unrelated to TSN* | |
| Ghazi et al. [84] | 2020 | | *Smart City related survey article, unrelated to TSN* | |
| Kirimtat et al. [121] | 2020 | | *Smart City related survey article, unrelated to TSN* | |
| Amodei et al. [5] | 2024 | | *Smart City related survey article, unrelated to TSN* | |
| Jagatheesaperumal et al. [115] | 2024 | | *Smart City related survey article, unrelated to TSN* | |
| Lopes et al. [140] | 2025 | Full tutorial | Smart City, Industry, Automotive, Avionics, Railway, and Power Grid | *none* |
| This article | 2025 | Only mention | Smart City, Industry, Automotive, Avionics, and Power Grid | 5G, 6G, and Wi-Fi |

# 3 | STATE-OF-THE-ART IN SYNCHRONIZATION COMPONENTS

When one mentions the concept of time within a network, one of two possibilities is questioned: is one trying to perform actions the best as possible regarding parameters such as latencies or jitter parameters, or does all the cooperating nodes in a communication have their clocks precisely synchronized between them or against a primary entity. As we are mentioning in the context of TSN, the second option is the one worth noticing, which is the reason why efforts were made, from 2011, to create a proper TSN profile for the usage of the precision time protocol standard: the **generalized precision time protocol (gPTP)**.

## 3.1 | Generalized Precision Time Protocol (gPTP)

The concept of acting "right now" became a reality for the TSN-based networks [218]. With a default synchronization time of eight times per second, gPTP can determine the worst-case time accuracy ahead of time, which can only be done by placing requirements of hardware timestamping for switches to participate in the protocol, measuring, and propagating the delays. In some cases, only a firmware update is required. Still, independently of this, the gPTP profile allows PTP to be run under an Ethernet context and extends its usage and configuration to IEEE 802.11 links. This increases the difficulties of its design due to the channel access latency, substantial jitter in frame transfers, topology changes, and features such as frame aggregation. These are some of the issues the scientific community is still trying to tackle, mostly ultimately with the attempt of launching TSN concept within the advent of 5G-and-beyond cellular networks [41, 24, 27, 200].

The ability to synchronize multiple network nodes within a proper and unique timing domain allows the scientific community and the industry to advance research and develop projects where PTP might enable the ordering of events, the distribution of data or coordinated computation, or even sensing data for control, cyber-physical, or vehicular-to-everything systems, among other applications and challenges [218]. More specifically, within TSN quests, this synchronization is vital for functions related to audio and video (AV) applications (which prominently work with isochronous streams of data, recalling the AVB profile of TSN), for applications of the time-aware shaper defined in IEEE 802.1Qbv standard [170].

### Synchronization Enhancements

As mentioned, the gPTP was packed with features allowing it to work under wire-bound and wireless scenarios. Still, issues arise when crossing these scenarios simultaneously, as in a hybrid-context network, or when specific nodes within the same network do not participate in a PTP exchange [76]. As shown by Baniabdelghany et al., in 2020 [21], different delays need to be taken into consideration in clock synchronization, as of deterministic delays (residence times of hardware-processing and propagation delays in both transmission and reception of time frames) and random delays (of channel access and transmission jitter). Some authors, such as Lam et al. [129], considered first to attenuate the clock drift with the frequency of the execution of the PTP procedures; secondly, they tried to tackle the asymmetric delays caused by the non-balanced hardware processing timings on both down- and uplink. The work of Baniabdelghany et al., moreover, considers variable surrounding conditions such as traffic congestion and link loss, focusing more on measuring the clock drift and the path delay. The adjustment of the frequency with which the PTP procedures are applied, and messages are exchanged is considered in work by Zhu et al. [255], which proposed a method to fix the timing synchronization triggered by clock drift estimations. More recently, Yu et al. [240] proposed attenuating and solving this issue by implementing a neural network model to analyze and predict time errors.

In 2021, in a work by Liu et al. [139], the authors studied the impact of link failures as a common factor affecting clock synchronization, proposing adding redundant ports to deal with sudden link drops while keeping energy consumption minimal.

Looking over the timing issues when using the time-aware shaper (TAS), problems arise when schedules are placed and the reference clock is lost. At this moment, individual clocks in each node participating in the same schedule or transmission start drifting apart until a new clock is elected as the new time transmitter clock[†]. Craciunas et al., in 2021 [50], addressed this problem, which may lead to a loss of determinism in the expected transmission time patterns programmed in the gate control list of TAS.

## Integration with other network technologies

Considering industry as one of the main use-case scenarios for implementing TSN solutions, some works were performed to allow the integration of the recent and more flexible TSN capabilities with the old and primarily proprietary solutions already deployed. In the work of Mateu et al., in [151], the authors mention and study the time synchronization requirements for integrating TSN techniques and mechanisms with EtherCAT as a legacy network in an automation industry. They have proposed combining the different clock synchronization mechanisms for precise agreement.

Sharing the same effort of attempting to integrate the TSN set of tools with other technologies, many research groups are trying to develop ways of integrating the standardized Ethernet-based protocols and mechanisms into the emerging 5G cellular network contexts [200, 117]. Nasrallah et al., in [163], started to survey studies targeting the support of ultra-low latency (ULL) in 5G networks in interaction with devices from wireless devices via access, backhaul, and core networks. Godor et al. [85] also surveyed the state-of-the-art in this integration in 2020, which conceptually bridged the time synchronization of the 5G system with the TSN domain synchronization procedures. Atiq et al. [17] state that a 5G integration with TSN can be made possible timely-speaking even with multiple TSN time domains, following the 3GPP Release-17 [1]. In the downlink TSN time synchronization, the grand-master (GM) clock is located at the network-side TSN translator (NW-TT) end (at the user-plane function (UPF)), and, on the contrary, the uplink TSN time synchronization has the TSN GM clock attached to the device at the device-side TSN translator (DS-TT) side (at the user equipment (UE)) [85, 96, 234]. Notwithstanding, efforts also exist to bring TSN developments into 802.11-based networks [18].

## Configuration ease and Hot Standby mode

Starting in 2021, a proposed amendment to gPTP was made by the IEEE 802.1 task group [193]. This amendment, denoted by IEEE P802.1ASdm, specifies protocols, procedures, and managed objects for the usage of a hot standby mode in which the best time transmitter clock algorithm (BMCA)[‡] is not used and all gPTP port are statically configured. In such a context, two grandmaster nodes coexist (a primary and a secondary), each creating its own PTP time domain, yet synchronized between them. This means all the handovers between both domains must be seamless [90]. This standard amendment is helpful in critical applications, for instance, where a communication outage could have negative repercussions.

Although being complemented by the concept of failure mitigation, the hot standby proposal does not define the procedures to handle such a function. According to the applicable use case, this feature is left unspecified for specific TSN profiles. Moreover, efforts are also being made regarding procedures for the configuration of these settings via the description of these characteristics for static determination of the grandmaster nodes and time domains through standardized YANG schemes, as proposed in another proposed amendment of IEEE P802.1ASdn [193].

Summarizing all these works found in table 2, it is possible to see the complete list of articles sorted by category relative to PTP in TSN.

[‡] In 2021, a new proposal was initiated in the IEEE 802.1 task group to create an alternative and inclusive terminology for the concept names used in the gPTP standard (IEEE P802.1ASdr). As a result of this amendment, the terms of *master clock* and *slave clock* are now to be mentioned as *time transmitter clock* and *time receiver clock*, respectively. The terminology of *grandmaster clock* remains unchanged, but the BMCA algorithm also changes to BTCA, as in best *time transmitter* clock algorithm.

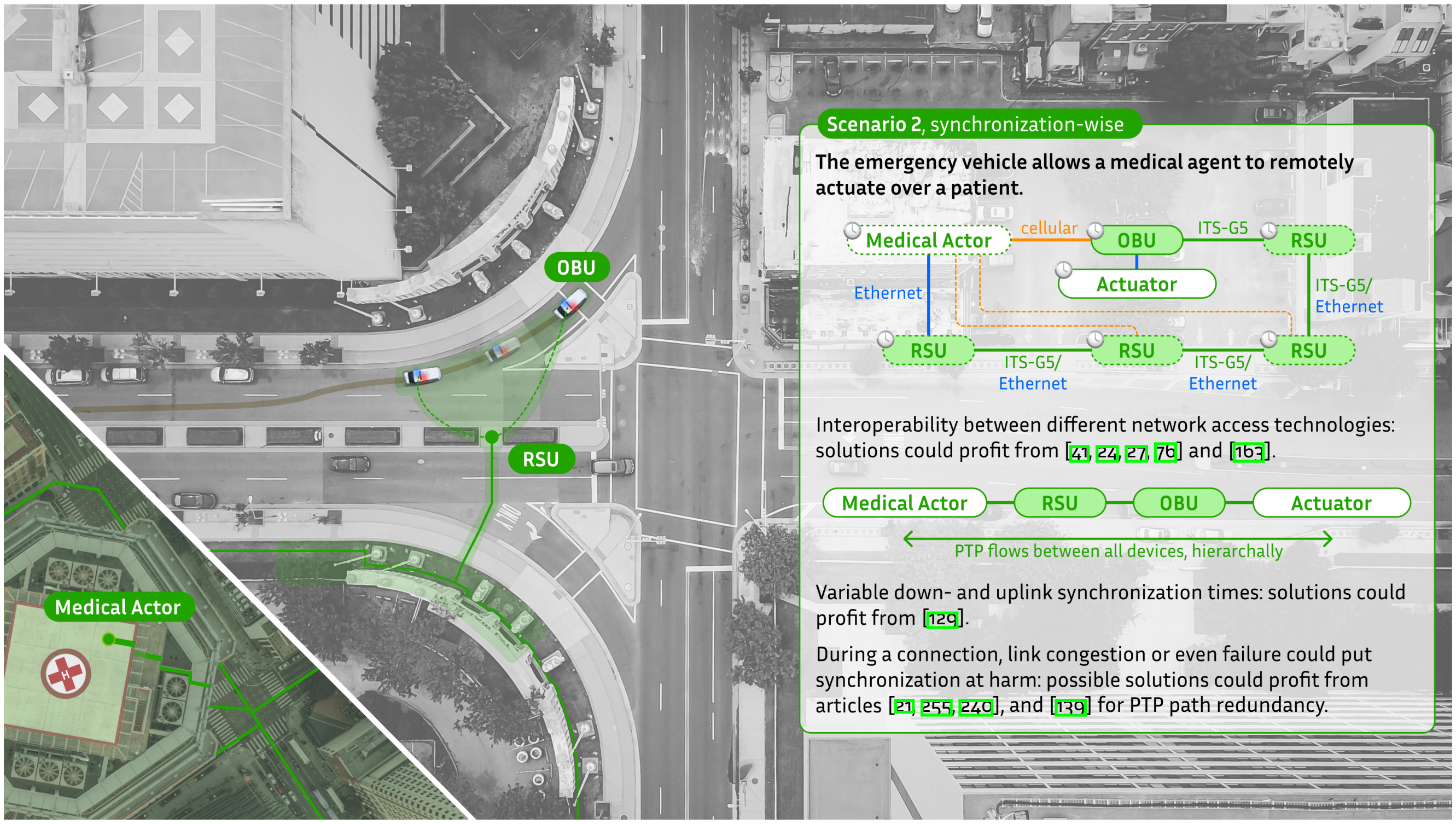

**FIGURE 9** Ambulance on a mission with reserved lane use case scenario, in a synchronization perspective

**TABLE 2** Summary of all collected related work to gPTP in TSN

| **1. Overall studies, context, and conceptualization** | **3. Interaction with non-TSN technologies** |
|---|---|
| [218], [41], [24], [170], [193], [204], [205], [27], [90], [147], [177], [5], [138], [160], [248] | [151], [117], [243] |
| **2. Synchronization issues** | **4. 5G-and-beyond interaction** |
| [76], [21], [129], [255], [240], [139], [50] | [163], [85], [17], [96], [234], [200], [5], [117], [239], [243] |

## 3.2 | Discussion

Looking over the current state-of-the-art concerning synchronization, it is quite clear that long are the advancements already taken by the scientific community. The gPTP standard already has profiles defined for some of TSN's most common use case scenarios (such as automotive, industry, and avionics).

In a smart city use case context, a specific profile does not currently exist. However, one can envision how such a system might operate by considering the smart city context as an encompassing set of all known TSN use cases. To illustrate this concept further, figure 9 provides a detailed view of the initial use case scenario highlighted in our motivation. To revisit the idea, this scenario involves an ambulance enabling a remote medical professional to intervene directly with a patient en route to a hospital.

In such a scenario, multiple network access technologies are utilized. However, TSN primarily supports Ethernet as its extension. It is imperative to ensure timely synchronization of all nodes involved in this scenario to avoid jeopardizing the life-assisting procedures being performed for the patient. Bad synchronization latency issues could disrupt commands from medical professionals reaching actuators or cause delays in transmitting vital signals to remote medical professionals. To address this concern, gPTP is employed within TSN. A defined hierarchy determines the node responsible for the grandmaster role, and PTP packets flow through the paths connecting all nodes.

Regarding the current accessibility in the implementation of this standard, there are plenty of works already designed and built to openly work in environments such as, for instance, Linux, by using the `linuxptp` package defined mainly by Richard Cochran (starting in [48]) with the help of contributors from industry-critical stakeholders such as Intel's and AudioScience's, among others.

Nevertheless, there is still work regarding time synchronization, as TSN designs and deployments consider the possibility of having more than one gPTP domain. Currently, there is no standard method to transform the synchronized times of two or more domains into one synchronized time without using the primary best time transmitter clock algorithm or simply mapping time references between domains. As mentioned before, there is an ongoing project of hot standby (IEEE P802.1ASdm) to tackle this question, which could potentially be a relevant solution in the context of a smart city as grandmaster nodes could be statically selected on dedicated network locations; the infrastructure could profit from redundancy factors regarding the number of PTP domains and clock sources; the multiple PTP domains allow redundant clock sources in case of synchronization errors or cyber-attacks to grandmaster nodes.

Moreover, work is also in progress regarding maintaining a simultaneous time reference in wire-bound and wireless connections, more precisely in 5G-and-beyond deployments, in integration with other wire-bound TSN topologies, and also regarding the time source over which all network elements of the smart city infrastructure work on. In work by Martins et al., [150], a hierarchical view for time synchronization within the scope of a smart city infrastructure was designed and implemented, distributing a clock source time via PTP, capturing pulse-per-second signals from GPS receivers in mobile nodes.

At the time of writing, these synchronization methods are not yet the primary method for Linux implementations due to hardware requirements. Precision hardware must be used with the ability of timestamp packets and have a PTP reference hardware clock implemented into the used NIC. Despite this not yet being the case, we believe future hardware will implement such components into their architectures, making PTP more common and standard regarding the alternatives in timely synchronizing machines in a network.

Another aspect discussed in figure 9 is the potential risk of link failure or congestion. These factors can significantly disrupt PTP messages, affecting the periodicity and precision of time synchronization. The figure indicates that existing research addresses this concern specifically for gPTP. Additionally, there are broader solutions beyond synchronization alone, which we elaborate on in the following subsection.

# 4 | STATE-OF-THE-ART IN RELIABILITY COMPONENTS

Regarding the topic of reliability, as mentioned in the subsection before, TSN comes packed, until the date of writing, with three main standards and mechanisms for implementing robustness and other performance metrics and guarantees in the deployments: the frame replication and elimination for reliability (FRER), the per-stream filtering and policing (PSFP), and the path control and reservation (PCR).

## 4.1 | Frame Replication and Elimination for Reliability (FRER)

Keeping such an order as a reference, let us start with **FRER** standard. Such a standard, described as such for the first time in 2017 [110], was already being discussed in the TSN workgroup in 2014, as Kehrer et al. mentions in [118]. Here, frames are duplicated and transmitted over disjoint paths until an intermediate node where duplicates are removed to ensure a TSN stream is fault-resistant.

This logic has been simulated in some works, such as an integration in the TSimNet tool [12] or the works of Heise et al. in [100]. It was tested in a topology where a data stream was sent from a source to a sink node, over which cross-traffic was injected in parallel. The results were somewhat satisfactory regarding FRER, as they consist of multiple stages in the relay unit. Once the frames are reunited into a single stream, this one is subject to cumulative recovery procedures. This is not to be understood as if this mechanism is ineffective, but not to be used exclusively, as other reliability features can ease the overhead caused by these procedures [215, 128]. Examples of this same conclusion are the works in [38], in which their developments are not using FRER but did consider evolving to its implementation.

### Spatial Redundancy

As stated initially in the FRER standard, these works reflect the characteristic of **spatial redundancy**. Even with its mentioned drawbacks, other authors tried to complement its logic with procedures that help reduce some overheads in the processes of FRER by implementing heuristics or better considering the environment where such a mechanism is to be applied [206, 56]. In 2021, in work by Syed et al. [220], the authors developed and proposed four heuristics to support FRER in an internal vehicle network use case scenario to choose better which paths are supposed to be used to transmit replicated frames towards a listener. In the alternative, Sambo et al. [197] proposed the application of a finite-state machine able to proactively produce redundant frames through different paths without causing harm to the bandwidth with unnecessary transmissions.

In an attempt to study such an issue in more detail, Atallah et al. [14] started by considering audio/video traffic with temporal jitter requirements of a maximum of 2 ms in more than seven hops with stream replicas. By constantly estimating the meantime to detect an error, the reliability of such transmissions was evaluated with the help of integer linear programming (ILP), concluding that it was possible to find solutions up to 50% less bandwidth usage than in most cases. This work was then expanded, in [13, 16], to other types of traffic, where the authors studied a larger routing mechanism with greedy heuristics, validating the same results and featuring scalability issues.

### Temporal Redundancy

Even though the main concerns of the usage of FRER are something to focus our attention on, some other authors preferred not to deal with their possible occurrence but to avoid them by mixing them with another concept of **temporal redundancy**, in which, by time, frames are replicated, ideally, only if required to.

Starting with a design of such a concept, Álvarez et al. in [256] planned the architecture of a mechanism able to deliver several frame replicas to be produced in a given context, also choosing which data is supposed to be subject to replication and also tackling with the possibility of creating and then handling errors in these procedures. Two years later, the same authors carried simulations in which they already had a plausible implementation that also used principles of spatial redundancy, calling such a perspective a proactive transmission of replicated redundancy to tolerate temporary link failures [257]. Also, in 2019, in the works of Dobrin et al. in [59], temporal redundancy is studied by applying such criteria to TSN schedules.

### Single-event upsets

During the transmission of a stream and the execution of several components and nodes participating in a network session, there is always the possibility of latent hazards, even in fully set TSN schedules. Following the aforementioned work by Atallah et al., the same authors in 2019 [15] have produced errors in transmissions' schedules to determine their weakest point and to properly test FRER's reliability features. They compared the robustness of either a single path or multiple paths while manipulating the schedule information.

These errors, being injected in the latter work of Atallah et al., are supposed to represent errors that can even occur for electromagnetic interference issues during transmissions, whose concept is usually designated by single-event upset (SEU) errors. In other works, this time by the hand of Feng et al. [73, 74], the authors attempt to apply reliability policies such as FRER and other less bandwidth-abusive own alternatives. Given this scenario, they propose a fault-tolerant algorithm based on schedule reservations in cooperation with other TSN mechanisms such as TAS.

### Link failures and alternatives to FRER

The scientific community also tackled the issue of fault detection regarding the usage of mechanisms of artificial intelligence and machine learning techniques. Some frameworks have been designed and put into validation procedures to accelerate scheduling updates according to the detection and presence of plausible failures [58, 53, 90].

Alternatively, Ojewale et al., in 2020 [169], proposed a proximate approach to FRER, not by exclusively applying a replication strategy, but by looking to network-level metrics, load-balance paths to where streams are being called to follow to reach their destination, avoiding network congestion scenarios.

In table 3, one can find the complete list of articles found, sorted by category relative to FRER in TSN.

**TABLE 3** Summary of all collected related work to FRER in TSN

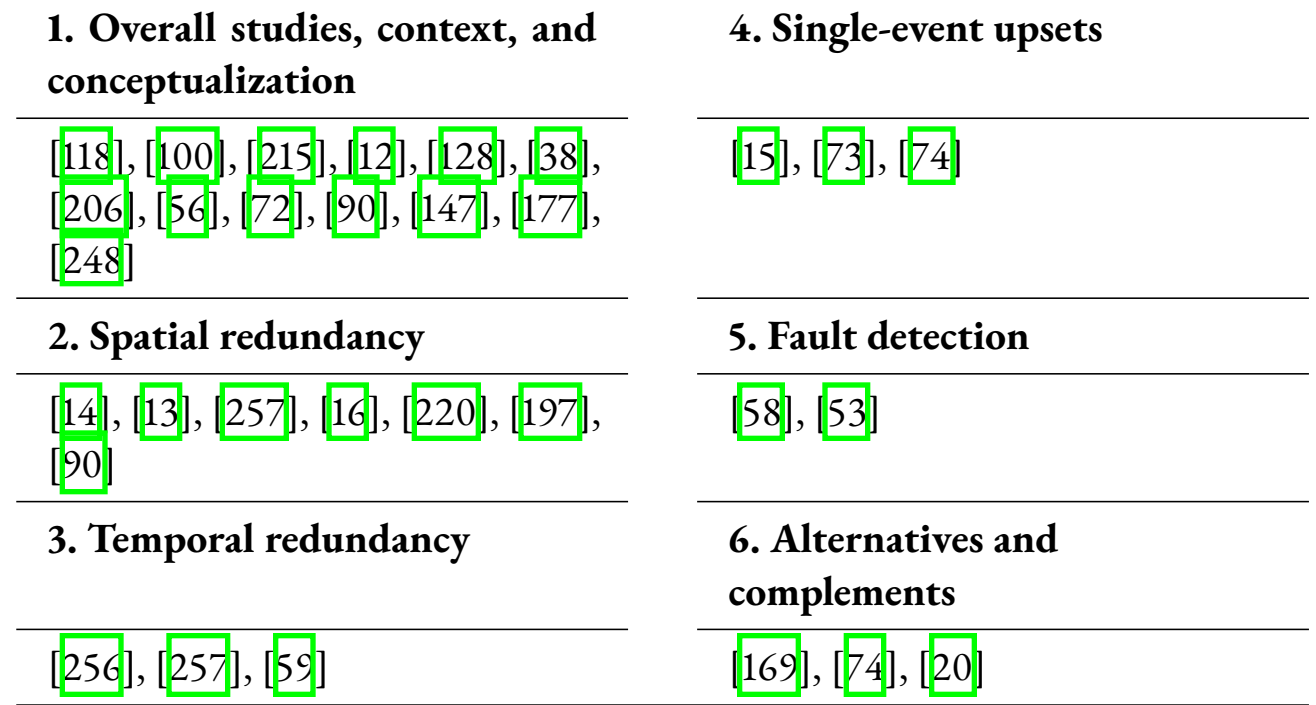

| **1. Overall studies, context, and conceptualization** | **4. Single-event upsets** |
|---|---|
| [118], [100], [215], [12], [128], [38], [206], [56], [72], [90], [147], [177], [248] | [15], [73], [74] |
| **2. Spatial redundancy** | **5. Fault detection** |
| [14], [13], [257], [16], [220], [197], [90] | [58], [53] |
| **3. Temporal redundancy** | **6. Alternatives and complements** |
| [256], [257], [59] | [169], [74], [20] |

## 4.2 | Per-Stream Filtering and Policing (PSFP)

Following the order we used to introduce the reliability concepts in the subsection before, the second standard in TSN to mention is the **PSFP**, defined in the IEEE 802.1Qci [109]. Our research has shown that this standard is still not widely used in the scientific community. Still, some references to it can be found in the works of Heise et al. [100] (where the concept is still being described as a draft, in 2016); Ashjaei et al. [12] (where the authors mention advantages in its usage such as to improve security, blocking a traffic source when unforeseen or not compliant traffic is detected); Deng et al. [56] (where the concept is also pointed out as a measure for security in the TSN toolbox of protocols and mechanisms); and in the technical report by Boyer et al. [32] (where a more algebraic analysis provides us a handful of base considerations for when using PSFP).

Despite the lower number of articles than the standard before, all the found works are included in table 4.

**TABLE 4** Summary of all collected related work to PSFP in TSN

| **1. Overall studies, context, and conceptualization** |
|---|
| [100], [12], [56], [32], [90], [147], [248] |

**TABLE 5** Summary of all collected related work to PCR in TSN

| **1. Overall studies, context, and conceptualization** |
|---|
| [118], [56], [147], [177], [248] |

## 4.3 | Path Control and Reservation (PCR)

Going back a while, and as initially designed in [110], FRER handles spatial redundant replicas as they are transmitted in a network in parallel on disjoint paths. Simultaneously, another TSN standard can help the first to make disjoint paths available towards a given end, being the **path control and reservation (PCR)** [108].

Along the collected works obtained as a result of our research, the usage of PCR is mainly done as a companion mechanism for FRER since it can provide it with the disjoint paths required for FRER to work as expected. Despite not being a widely implemented standard in the scientific community, the standard first appeared when describing such a seamless integration with FRER in 2014, still as a draft and future TSN integration in the work of Kehrer et al. [118]. Later, the work was found as described along FRER in a survey by Deng et al. [56] as a standard for link redundancy combining the usages of the intermediate station to intermediate station (IS-IS) routing protocol and of shortest path routing (SPR) method.

Similar to the standard before, all the works relative to PCR are presented in table 5.

## 4.4 | Discussion

When studying and further deploying a new network concept onto a set of machines or solving a given high-standards problem, it is quite easy to start with the solution *per se* first and then follow up with criteria to allow such a development to keep up along the time, providing performance guarantees such as of availability, robustness, and reliability.

Since it was first thought of and included as a primary component of TSN, the reliability features of TSN are mostly done regarding standardization but still in progress regarding both adoption and implementation by the scientific community. Finding some implementations stated by some vendors already installed in their equipment, mechanisms such as FRER, PSFP, and PCR are some examples of reliability standards that still lack existence in a Linux system in open-source for a global community to adopt in their deployments. The industrial-based community has standards that can achieve similar goals to what FRER targets, for instance, such as IEC's parallel redundancy protocol (PRP) or high-availability seamless redundancy (HSR). Still, they both have strict requirements to retrieve their full potential: PRP needs a parallel network so it can enable the frame duplicates, which adds to the implementation costs of adding more infrastructure components or rules, and HSR can only be set in the context of a ring network topology to run in its complete set of features [55]. TSN's logic behind FRER does not require any of the above-mentioned requirements, and it is thought to be a unified paradigm for reliability in these types of networks.

Moreover, as mentioned, some mechanisms are still being discussed to transition from their current draft shape to a standard. From all the standards stated here, FRER has the most development done and is hoped to be the first to have shape onto a Linux package in the near future. This would bring a common way to ensure redundancy and delivery guarantees in Linux-integrated TSN deployments.

As we delve into our smart city use case, the significance of reliability features cannot be overstated. Illustrated in figure 10, and focusing on the third use case scenario outlined in our motivation, wherein a building experiences a gas leak, these features play a crucial role. They empower the system to capitalize on redundancy levels, ensuring that alternative paths are readily available to sustain uninterrupted service in case of a critical link failure or congestion. In this scenario, redundancies can be established via Ethernet through RSUs. This approach guarantees seamless delivery and processing of critical data related to the gas leak, enabling actions like shutting off nearby gas switches. Furthermore, it extends to data transmission concerning the positions and maneuvers of emergency and autonomous vehicles. This cooperative network can also encompass non-autonomous vehicles, providing vital information to drivers, for instance, through traffic lights to clear obstacles from the path of emergency vehicles.

As seen in the selected surveys on smart cities from table 1, authors mention that redundancy is one of the key features an infrastructure could profit from as a positive attempt to outperform the availability metrics for the network functions provided by the city. This set of reliability protocols and mechanisms is fundamental to allow environments of smart mobility (transportation features that can enable emergency scenarios), smart environment (where network resource management could be performed with sustainability in mind), and smart living (where adjacent network and sensing functions to buildings and other infrastructures are more reliably transmitted from node to node) [121, 84]. The heterogeneity of the network infrastructure nodes also builds up the requirement for reliability features since the management and each node handles the way procedures differ according to their function [60].

# 5 | STATE-OF-THE-ART IN LATENCY COMPONENTS

The most common pillar of TSN to be mentioned by anyone working within the scope of time-sensitive data exchanges is latency.

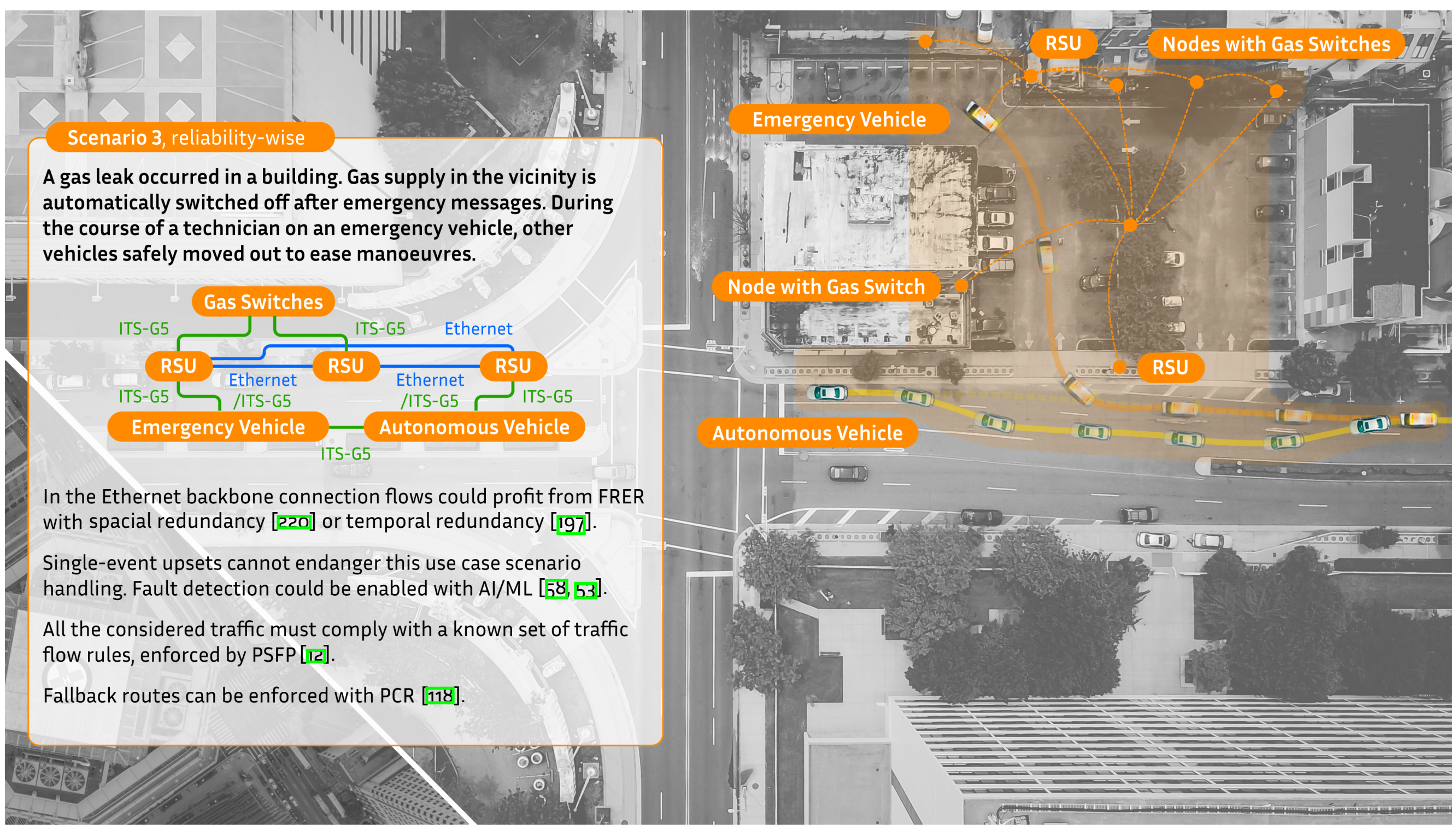

**FIGURE 10** Infrastructure safety provisioning after a gas leakage use case scenario, in a reliability perspective

This category of standards and mechanisms is where TSN has tools for transmissions to be done at the right time rather than as quickly and as best as possible. The work in the literature on the different types of shapers to decrease latency is presented in the following sections.

## 5.1 | Time-aware Shaper (TAS)

Mentioning latency standards and mechanisms for TSN, the time-aware shaper (TAS) must be the first and more typical example of a TSN standard used in the scientific community. Currently also addressed by its standard IEEE 802.1Qbv, the first works mentioning this shaper were developed by Thiele et al. and by Thangamuthu et al., in 2015 [224, 221], respectively, in which a formal worst-case timing analysis was performed for TAS.

In 2016, Craciunas et al. [49, 51] stated that the transmission of hard real-time traffic within a fully-time-synchronized context is almost equivalent to scheduling features in TAS, allowing time set procedures of opening and closing gates at an egress port of bridges. Along with other authors [198, 52, 116], there are workpieces that, directly mentioning frame-level scheduling being an NP-complete problem, study general formulations by the usage of ILP and other techniques, or advance to simulations using algorithms controlling better scheduling procedures or applying heuristics to them. In [116], Jin et al. proposed heuristics to combine both scheduling tables and control packet injections to refrain from scheduling conflicts.

### TAS serving as a primary standard

Considered as a TSN standard able to enable a cost-effective integration of safety-critical and real-time distributed applications in domains where determinism is a requirement, such as aerospace, automotive, and industrial domains [51, 212, 211, 135, 117, 199], several authors validated such a statement for TAS with simulated tasks in the automotive scenario [175, 176, 122, 177, 6], in the industry [198], as well as in space and avionics [15].

Moreover, and although this standard may be the most complete and studied one of the set of TSN tools, the work on and with TAS is not over. For instance, efforts are being made to make time-sensitive traffic in contexts of Industry 4.0 and 5G networks a common reality [229, 12, 200]. Chuang et al., in [45], posed the question that one of the most critical issues for advancements in TAS was that, although there are some existing forwarding approaches by configuring gate control lists, as they are offline algorithms, they are not suitable for future industrial automation requiring dynamic reconfiguration procedures.

Notwithstanding, several works aim to use (or already use) TAS as a proper standard to obtain a time-critical solution or, as a matter of example, to get part of TSN benefits.

In the industry context, real-time communications are attempted to be performed with ultra-low latencies, not disregarding non-critical data flows from being transmitted in the same medium but with different priority definitions. Within this paradigm, authors try to prove and use TAS as a candidate for shaping how and when critical frames are transmitted throughout a network [9, 73, 97, 38, 29, 77, 81, 74].

Another usual application for TSN is automotive, where the scientific community has already created several approaches to use standards such as the time-aware shaper. Although most of them do not use proper equipment other than running simulations, an automotive profile for TSN has been addressed with the use of TAS in several works [254, 28].

At a different scale, compared with automotive, some work was found to study or implement TAS techniques onto train and avionics applications, as well as some unmanned aerial vehicles (UAVs) [152, 22].

## Considering routing paths

Earlier works on the TAS did not consider the possibility of transmissions being forwarded via paths other than the pre-known shortest path, as evaluated by most switching software. Considering additional routing paths, whilst the easiest or more intuitive way would be to select the available paths with the minimum number of hops, this alternative has some issues: allowing traffic to be routed to a preferable path with a minimum number of hops can lead critical-traffic to suffer consequences due to bottlenecks [206, 235]. Furthermore, as frame-level scheduling is an NP-complete problem, most solutions are fixed routes in a commonly reduced solution space. For this reason, researchers are currently developing scheduling solutions in a more extensive solution space after considering sets of flows by joining the problems of routing and scheduling simultaneously. Even so, issues arise as one wants a valuable solution in time to be installed before time-constrained traffic is forwarded through TSN bridges: the lack of determinism in such scenarios could arise from errors in the synchronization procedures, inefficient temporal or spatial isolation; or even inconsistent network configuration updates [42].

In the search for an optimal solution, a first set of approaches to these problems was made using heuristics. Dürr et al., in 2016 [62] used tabu search to efficiently reduce the size of the solution space in computing schedules and reduce the number of guard bands in the schedule. Other authors followed similar approaches, such as in [78] and [237], but other heuristics were used such as greedy approaches by Laursen et al. [130], as well as other authors in [181, 79, 14, 174, 241, 116, 242, 190, 136, 81, 230, 137].

Alternatively, a mixture of techniques can be found to tackle these problems. Pop et al. [181] proposed a two-part solution. First, it applies a greedy randomized adaptive search procedure-based heuristic for the routing problem and then applies ILP to solve the scheduling problem. As mentioned earlier, some other authors have decided to apply ILP to solve both routing and scheduling problems jointly [201, 214, 164, 165, 3, 15, 65, 202, 241, 104, 137].

Other techniques, such as satisfiability modulo theory (SMT), were used to formulate the joint routing and scheduling problem. In [142], Mahfouzi et al. used SMT to address the routing and schedule synthesis issue and then provided heuristics to improve its efficiency based on route subsets and time slices. By similar approaches, other authors in the scientific community followed their developments [170, 143, 198]. In such a list, Santos et al. [198] has made openly available its framework for generating TSN schedules.

Opting for better formalizing the problems, some authors proposed their developments by studying them under a network calculus framework [207, 208]. Using completely different reasoning, a solution using source routing is also proposed [127], and others use more versatile approaches by using machine learning procedures [144, 145].

## Guard Bands usage for mixed-criticality

Due to the nature of TSN, various traffic types can coexist in a network; critical and non-critical traffic can be simultaneously requested to be transmitted along the connections. In such a scenario, which one can call mixed-criticality, using a guard band (GB) can be helpful to ensure the beginning of a scheduled traffic section of the packet flows.

Some of the works mentioned earlier already refer to some considerations on mixed-criticality while solving the problems for the routing and the scheduling of time-triggered traffic [62, 214, 181, 130, 79, 212, 15]. Still, they ignored lower priority real-time traffic (such as of AVB), which results in time-triggered configurations that can increase the worst-case delays of such decreased priority real-time traffic, making it non-schedulable [80]. In 2019, in a work by Heilmann et al. [99], the authors proposed the creation of a size-based queue strategy to improve the bandwidth utilization under the context of mixed-criticality, where TAS is used along with guard bands. This method was shown to reduce the amount of wasted bandwidth without causing any impact on the transmission of high-priority messages or requiring extensive changes to the network [99].

Alternatively, in the works of both Zhang et al. [246] and Cao et al. [40], traffic is re-categorized for different priorities to be coordinated in their transmission. In the first development, the authors categorize traffic into three distinct groups: high time-sensitive flows (such as motion control in industrial applications, where their critical delay and jitter requirements must meet the control performance guarantees), low time-sensitive flows (such as audio/video bridging systems), and best-effort flows. Following this procedure, the authors developed a parameter selection approach for choosing the proper cycle time and placed an injection time grouping algorithm to coordinate time-sensitive traffic. On the other hand, Cao et al. [40] did not ignore the possibility of best-effort traffic also getting the opportunity of being scheduled, as the critical traffic is organized according to subperiods, reducing the average scheduling intervals of such frames.

In a completely different approach, considering a taxonomically distinct critical scenario, Kim et al. [120] proposed the study of non-isochronous emergency traffic in time-sensitive networks. In this research work, the authors have described an enhancement to the time-aware shaper in which emergency traffic is immediately forwarded whilst real-time performance is guaranteed. Emergency traffic is considered non-isochronous since nobody can expect such traffic to occur while schedules are being estimated: they are always a possible event to occur.

Many articles study the effects of TAS components or the complete mechanism. In 2017, Kentis et al. in [119] studied the impacts of port congestion in the duration of the network-wide schedule. The conclusion gathered was that a congested port can make scheduling more complex, which allowed the authors to consider port congestion as a convenient metric to care about during a scheduling procedure, which they have shown to reduce up to 26% the duration of the gating schedule in multipath networks.

## Delay Analysis

The time-aware shaper in TSN has a set of characteristics that makes it hard to optimize in an online fashion. The many variables needed for consideration are substantial, such as runtime, schedulability, latency, resource utilization, jitter, scalability, schedule length, and optimality, among others [42].

Some works provide delay analysis over the mechanism of the time-aware shaper. Proceeding with a study on the effects of the gate control lists, Atallah et al. [15] analyzed the impact of SEUs being injected onto a TAS-enabled TSN network. This schedulability analysis can also be found in other works in which authors study the impact of: multiple independent periodic TSN talkers [7]; the best-effort traffic in TSN, as it can itself be (or not) scheduled [102]; the AVB traffic in TSN [153]; the response time of TSN comparing both scenarios of time-aware shaper usage, versus frame preemption [25, 86]; or of the global traffic being scheduled, but performed with the help of machine learning trained model whose effects can lead to a given probability of false positives for the programming of the GCLs [145].

Nasrallah et al. [163] estimate both minimum and maximum delay and packet loss levels, creating an approach to an average-case analysis of the TAS standard usage in TSN networks. At the same time, other authors thoroughly performed a worst-case delay analysis for TAS [213, 191, 254, 81].

There are some articles in which studies can advise researchers and development teams working under TSN scenarios to adjust and consider their choices in configuring TAS. These studies analyze the criteria one should have when choosing between stream-based or class-based TAS and frame preemption [101, 187, 10]. Moreover, other authors also study the possibilities of bringing such a TAS concept to different types of networks, such as 5G-and-beyond networks [228, 94, 200].

## Simulations and Implementations

In a more practical sense, simulation frameworks and implementations have started to appear in the scientific community since 2018. Nevertheless, most works mentioned to this point have their results or verifications done using a simulated environment or set of procedures.

In 2021, Turcanu et al. in [227] presented the potentials of mixing TSN wired networks and best-effort wireless networks for vehicle-to-vehicle (V2V) communications, showing first insights on synchronized in-vehicle networks schedules over a plain IEEE 802.11p connection using well-established network simulators such as of OMNet++[†], INET[‡], Plexe[§], and Veins[¶]. On the other hand, some authors, such as Polachan et al. in [179, 180], have designed and proposed an open-source discrete-event network simulator by the name of PYTSN containing models of TAS-enabled TSN bridges and tactile cyber-physical system terminals[#].

Regarding implementations, and as Seol et al. so evidently emphasize in their survey in [206], there is still "the need for better equipment" and, most of all, "a development environment for researchers to study fast-evolving TSN". There are already some TSN-supported devices. Still, most of them are commercial off-the-shelf (COTS) solutions, which allow the ease of configuration and use but are relatively limited in their functionalities and flexibility for researchers. On the other hand, there are already some development kits open for researchers, such as boards produced by enterprises such as Relyum's[‖] (a branch of SoCe[**]) [162] or InnoRoute's[††].

With Relyum's hardware, work is already found validating and experimenting with their implementation [47]. Transitioning from simulation to implementation, some experiments use extensions to Mininet [127]. Some developments have already been performed regarding integration with the Linux operating system. Oge et al., in [168], presents the design, implementation, and evaluation of TAS, a task which, as mentioned before when we described the state-of-the-art in PTP, requires hardware-level timestamping features to reach the best precision as possible: to tackle this issue, the authors used network interface cards equipped with an Intel's I210 chip.

† https://omnetpp.org
‡ https://inet.omnetpp.org
§ https://plexe.car2x.org
¶ https://veins.car2x.org
# https://github.com/kpol-iisc/PYTSN
‖ https://www.relyum.com/
** https://www.soc-e.com/
†† https://innoroute.com/

**TABLE 6** Summary of all collected related work to TAS in TSN

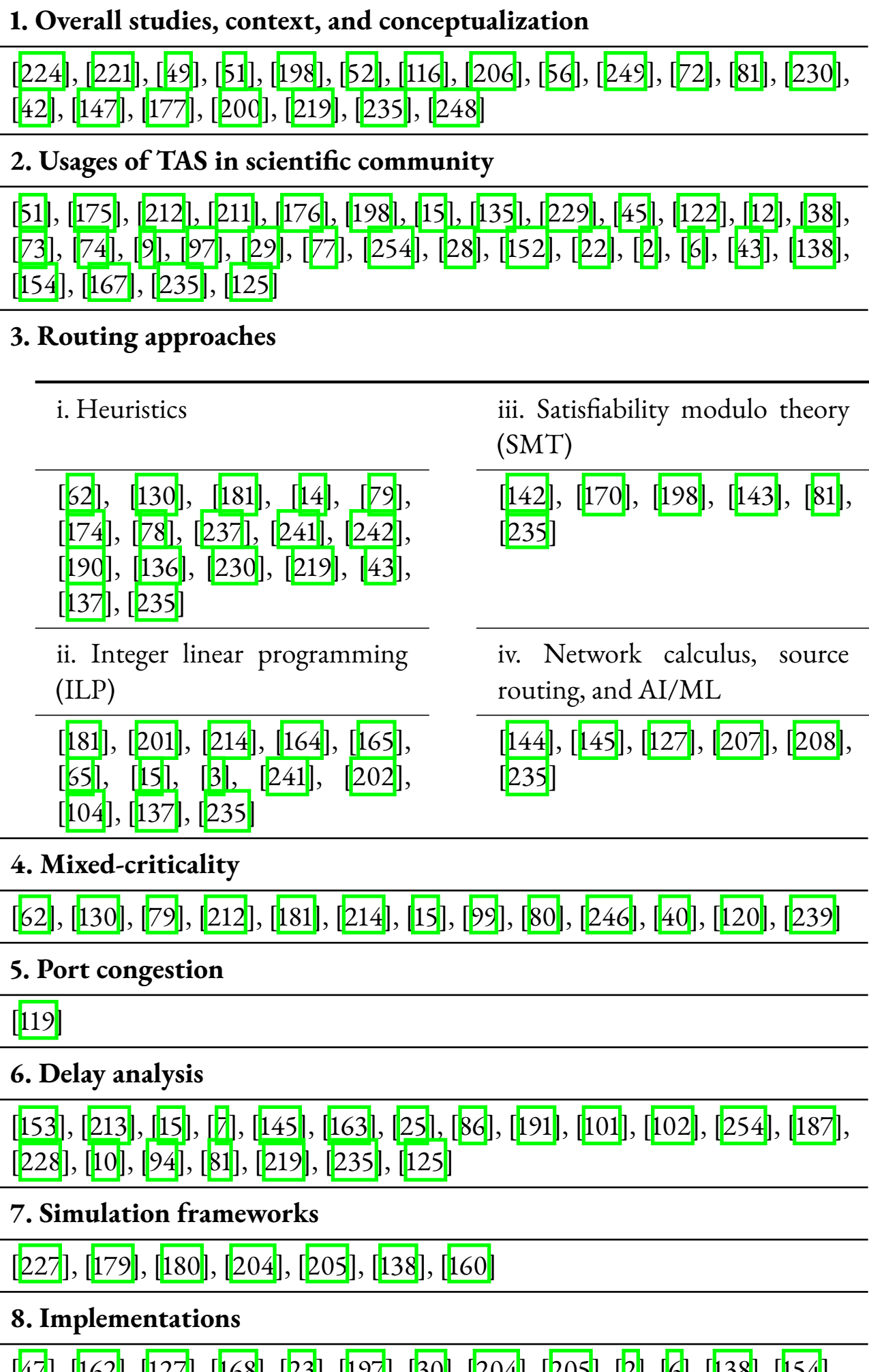

|  |  |
|---|---|
| **1. Overall studies, context, and conceptualization** | [224], [221], [49], [51], [198], [52], [116], [206], [56], [249], [72], [81], [230], [42], [147], [177], [200], [219], [235], [248] |
| **2. Usages of TAS in scientific community** | [51], [175], [212], [211], [176], [198], [15], [135], [229], [45], [122], [12], [38], [73], [74], [9], [97], [29], [77], [254], [28], [152], [22], [2], [6], [43], [138], [154], [167], [235], [125] |
| **3. Routing approaches** — i. Heuristics | [62], [130], [181], [14], [79], [174], [78], [237], [241], [242], [190], [136], [230], [219], [43], [137], [235] |
| **3. Routing approaches** — ii. Integer linear programming (ILP) | [181], [201], [214], [164], [165], [65], [15], [3], [241], [202], [104], [137], [235] |
| **3. Routing approaches** — iii. Satisfiability modulo theory (SMT) | [142], [170], [198], [143], [81], [235] |
| **3. Routing approaches** — iv. Network calculus, source routing, and AI/ML | [144], [145], [127], [207], [208], [235] |
| **4. Mixed-criticality** | [62], [130], [79], [212], [181], [214], [15], [99], [80], [246], [40], [120], [239] |
| **5. Port congestion** | [119] |
| **6. Delay analysis** | [153], [213], [15], [7], [145], [163], [25], [86], [191], [101], [102], [254], [187], [228], [10], [94], [81], [219], [235], [125] |
| **7. Simulation frameworks** | [227], [179], [180], [204], [205], [138], [160] |
| **8. Implementations** | [47], [162], [127], [168], [23], [197], [30], [204], [205], [2], [6], [138], [154] |

Other authors used the same chip [154], and Basumatary et al., in [23], proposed an implementation tested in a Linux kernel built with real-time concerns (with the `PREEMPT_RT` flag), and using the TAS scheduler configuration via the queueing discipline of TAPRIO[†], made available through the `tc` (*traffic control*) command. This very same usage of the TAPRIO queueing discipline was also found in the works of Sambo et al. [197], and their performance was evaluated in the study by Borda et al., in [30]. The usage of this queueing discipline is also being studied in virtualized environments [43]. Moreover, some authors evaluated TAS approaches on commercial off-the-shelf equipment and with open-source software [204, 205, 2].

[†] https://man7.org/linux/man-pages/man8/tc-taprio.8.html

To sum all the articles in categories relative to their relevance concerning TAS, table 6 has a full-length list of the selected works, comfortably sorted.

## 5.2 | Credit-based Shaper (CBS)

Alternatively to the time-aware shaper, the credit-based shaper algorithm allows traffic to be shaped in its transmission according to the bandwidth negotiated for a stream reservation while also controlling burstiness. The operation of CBS defines that credit is given to a given queue and increases and decreases at distinct rates when a frame is waiting for transmission or transmitting, respectively [211].

Many studies and developments using CBS are still from the AVB era [221, 224], whose concepts evolved and were transitioned to TSN [12, 56]. For this reason, following TAS, CBS is the second most applied concept of TSN when criteria of latency are to be applied in a time-sensitive network [4, 152, 244, 97, 28, 134].

### Delay Analysis

Similarly to TAS, the analysis of latency levels of critical traffic with CBS has actively been researched. Primarily, in 2017, He et al. [98] performed a study on the improvements of using a combination of CBS and TSN in comparison with more specific technologies such as avionics full-duplex switched Ethernet (AFDX), resulting in some configuration suggestions on the setting of CBS bandwidth. In the following years, the scientific community, already more open to the TSN technologies as they became a standard, studied which approaches to take by comparing several mechanisms working cooperatively: authors examined the impact of using TAS in comparison with CBS, or even with using frame preemption [187, 10].

Since CBS transitioned from AVB to TSN, the analysis of worst-case performance and response times have also been object of study in the scientific community [153, 250, 159, 158, 245, 66, 102, 34, 209].

### Experiments and Implementations

Some experiments were conducted to vary the tests and procedures with CBS. Boyer et al., in [33], studied the impact of changing the credit evolution rules in CBS, in this case, on freezing the credit evolution before gate closing. In another work, Fang et al. [67] developed a token-regulated credit-based shaper as an extension to the main CBS algorithm in searching for a more fine-grained quality of service distinction of the traffic.

Regarding implementation, some work has already been developed in Linux with the help of queue disciplines in Linux's `tc` software, as demonstrated by Borda et al. [30]. Similarly to TAS, Cid et al. [47] tested the implementation of CBS in the TSN-capable boards by Relyum.

A simple view of the set of collected articles can be seen in table 7, sorted by category.

**TABLE 7** Summary of all collected related work to CBS in TSN

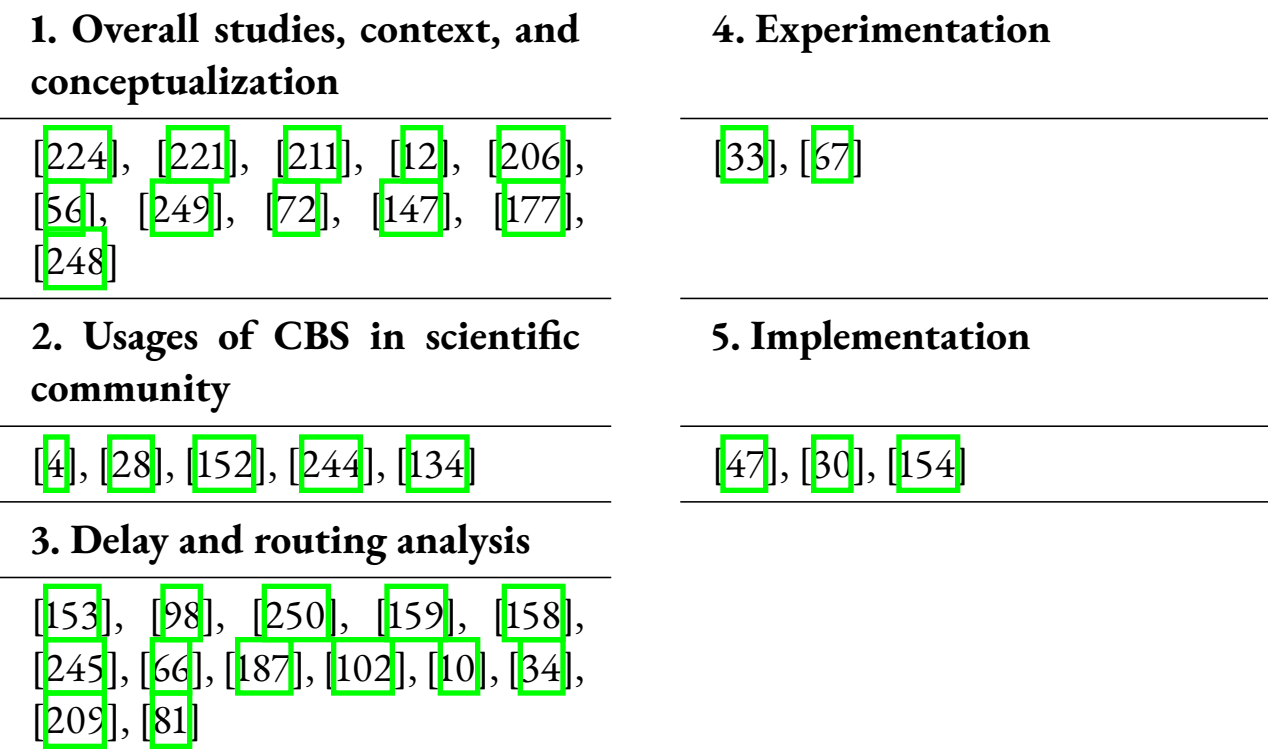

| **1. Overall studies, context, and conceptualization** | **4. Experimentation** |
|---|---|
| [224], [221], [211], [12], [206], [56], [249], [72], [147], [177], [248] | [33], [67] |
| **2. Usages of CBS in scientific community** | **5. Implementation** |
| [4], [28], [152], [244], [134] | [47], [30], [154] |
| **3. Delay and routing analysis** | |
| [153], [98], [250], [159], [158], [245], [66], [187], [102], [10], [34], [209], [81] | |

## 5.3 | Asynchronous Traffic Shaper (ATS)

Still describing the current state-of-the-art in TSN shapers, and as both time-aware and credit-based shapers were already referred, we are only missing now a literature review of the asynchronous traffic shaper (ATS). Firstly named the urgency-based scheduler (UBS) as described by Specht and Samii in [216], the ATS was designed to give low delay guarantees to traffic while absorbing jitter with the help of a pre-flow interleaved regulator, as they were deeply described in [217]. This type of shaper does not require a clock reference, in opposition to the latter ones, as mentioned by Zhou et al. in [253], showing that ATS can achieve effective traffic shaping and switching disregarding any synchronous mechanisms. This very same practice was tested in an industrial traffic flows scenario by Prados-Garzon et al. in [185] and by Mifdaoui et al. in [156].

Still under the name of the urgency-based shaper, the ATS was used by Garvriluț et al., in [82], to support the work in a fault-tolerant topology for TSN. Moreover, the most common use case scenario currently using ATS is the industry [225, 156] in which the usage of regulators is balanced and reviewed against latter shaper procedures.

### Delay Analysis

Through these works, the authors also addressed several analyses for the delay using various methods such as numerical analysis or simulations [159, 31, 88, 163, 103, 251, 66, 184, 46, 207, 187, 240, 12]. Nasrallah et al. in [163] have shown in their experiments that ATS can ensure the deterministic transmission of real-time streams, but it is not suitable for periodical critical traffic such as TAS, since it provides more fairness for all traffic types [56].

### Evolution towards 5G-and-beyond

Evolving in the methodologies to cope with the complexity of asynchronous flows, Prados-Garzon et al. [184] have implemented a reinforcement learning mechanism. The same authors, in 2021 [183], published a study on the adoption of asynchronous TSN for 5G back-hauling in which the ATS is seen as the main building block of asynchronous TSN.

Likewise to all the standards described before, in table 8 one can consult a summary of all the collected articles related with ATS, by category of study.

**TABLE 8** Summary of all collected related work to ATS in TSN

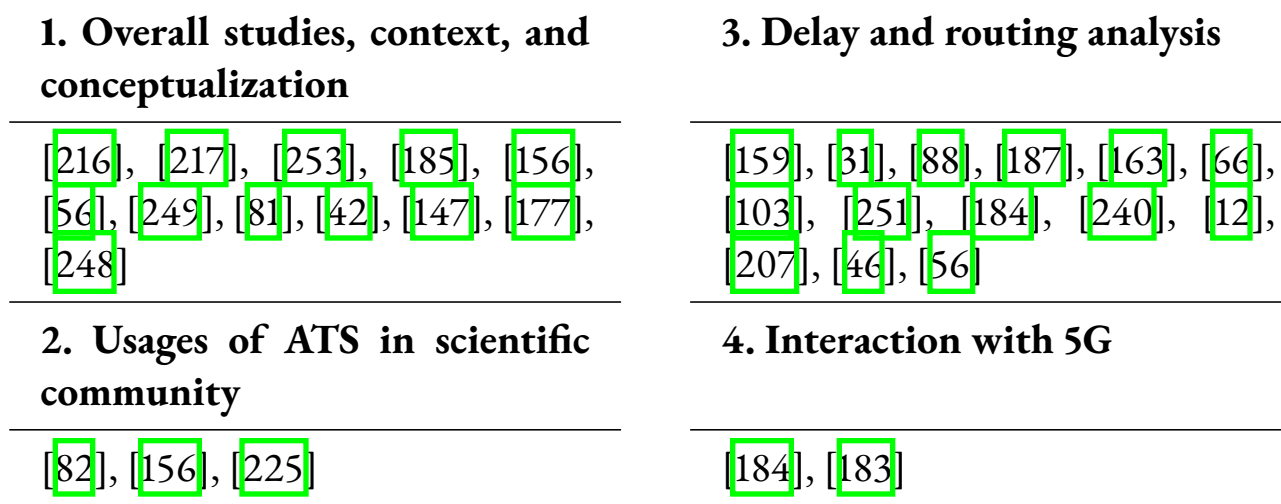

| **1. Overall studies, context, and conceptualization** | **3. Delay and routing analysis** |
|---|---|
| [216], [217], [253], [185], [156], [56], [249], [81], [42], [147], [177], [248] | [159], [31], [88], [187], [163], [66], [103], [251], [184], [240], [12], [207], [46], [56] |
| **2. Usages of ATS in scientific community** | **4. Interaction with 5G** |
| [82], [156], [225] | [184], [183] |

## 5.4 | Frame Preemption (FP)

Under the context of latency as one of four pillars of TSN, another mechanism arises rather than shaping mechanisms. One of them is frame preemption (FP), in which bridges are set to select traffic according to a given set of specifications, capable of categorizing it as express traffic. Despite this approach, frame preemption is also used to ensure the bounded worst-case end-to-end latency critical traffic [206, 72, 86, 250, 12].

The scientific community has tried to evolve this standard in the last decade since the AVB era. Lee et al., in 2016 [132], introduced a method of reducing the guard band by using FP, whose work continued as published in [133]. In the work by Zhou et al. [252], frame preemption is simulated, in which interspersed express traffic is evaluated. At the same time, the mechanism handles the preemption of the non-time-critical transmissions, minimizing the jitter. Other authors considered the near examples to it but regarding a set of use-case scenarios such as video traffic [124], industrial field control [212], or automotive [247].

**TABLE 9** Summary of all collected related work to FP in TSN

| **1. Overall studies, context, and conceptualization** |
| --- |
| [250], [86], [206], [12], [249], [72], [147], [177], [138], [248] |
| **2. Usages of FP in scientific community** |
| [252], [132], [124], [212], [133], [247], [245], [152], [9], [29], [28], [138] |
| **3. Delay and routing analysis** |
| [223], [132], [252], [133], [88], [25], [101], [86], [102], [11], [93], [206], [10] |

### Delay Analysis and Guard Band issues

As in the other mentioned TSN standards, regarding FP, the authors also performed theoretical and simulated analysis to the worst-case delay upper bound of its influence using various methods, such as numerically [223, 252, 132, 133, 11, 102, 10], by simulations [101, 25, 88, 93, 10], and by implementation [86].

Gogolev et al., in [86], implemented FP in a system-on-a-chip to test a six-hop daisy-chained bridges network with some non-synchronized end-devices. The authors concluded that frame preemption has a reasonable performance in the transmission of high-priority traffic in networks where time-aware shaper is unsuitable but disregards effects on low-priority traffic [206].

Ashjaei et al., in [11], identified a limitation in FP in which high-priority frames can experience significant blocking delays, consequently expanding their worst-case response times. Such a scenario would happen if, considering two preemptable flows (one with more priority than the other) are preempted as soon as an express frame is to be transmitted. If, during this preemption time, the preempted traffic is the lower priority one and then a higher one appears, the egress order is the same as ingress, not privileging the highest priority concerning the lower of the preemptable flows. To solve this issue, the authors proposed and developed a more efficient model for FP, allowing over 90% reduction in the maximum blocking delay, in which the priority within the set of preemptable flows is respected.

### Experiments

Lastly, plenty more articles showcase examples of FP usages in the scientific community developments and publications [29, 9, 28, 245]. Mauclair et al. [152] studied the advantages of adopting TSN standards in helicopters, in which FP plays a significant role in preempting non-critical transmissions in between the critical flows.

In table 9, a description of all the articles found is made, organized by categories.

## 5.5 | Cyclic Queueing and Forwarding (CQF)

Completing the set of latency-based TSN standards, there is still one standard left, which is a composition of PSFP and TAS [156]. This way, bridges are periodically synchronized in enqueueing/dequeueing frames with bounded latencies, depending only on the number of hops and cycle times [206].

### Experiments

There are some reported usages of this standard in the scientific community. Still, some authors have stated that some implementations suffer from removing the dependence on synchronous timing while providing a bounded delay [141]. Nonetheless, others use cyclic queueing and forwarding (CQF) to provide deterministic latency guarantees since, firstly, the sending and the receiving time slots of a packet on two adjacent bridges must be the same and, secondly, a packet received at a time slot must be sent at the next time slot in a bridge [238].

Another issue from the usage of CQF is related to the practicality of the standard in dynamic TSN application scenarios, where new scheduling algorithms are required. Quan et al. [188] proposed an online flow injection time scheduling algorithm to explore the deterministic ability of CQF by incrementally generating traffic schedules for new time-sensitive flows, adjusting the sending time on the network adapter based on network resource utilization.

### Delay Analysis and Enhancements

With the help of numerical and heuristic analysis methods, there is also work on the latency analysis of CQF, such as the ones developed by Yan et al. [237] and Zhang et al. [244].

Moreover, there is also work of enhancement of CQF in which the main objective is to solve the bandwidth, cycle, and queue mismatch problems in end-to-end scheduling, as developed and proposed by Huang et al., in [105] and [106].

Similarly to the standards before, in table 10 one can consult the complete list of articles found related with CQF.

**TABLE 10** Summary of all collected related work to CQF in TSN

| **1. Overall studies, context, and conceptualization** | **3. Delay and routing analysis** |
| --- | --- |
| [206], [156], [72], [106], [42], [147], [177], [248], [232] | [237], [244] |
| **2. Usages and issues** | **4. Enhancements** |
| [238], [188], [141] | [105], [106], [232] |

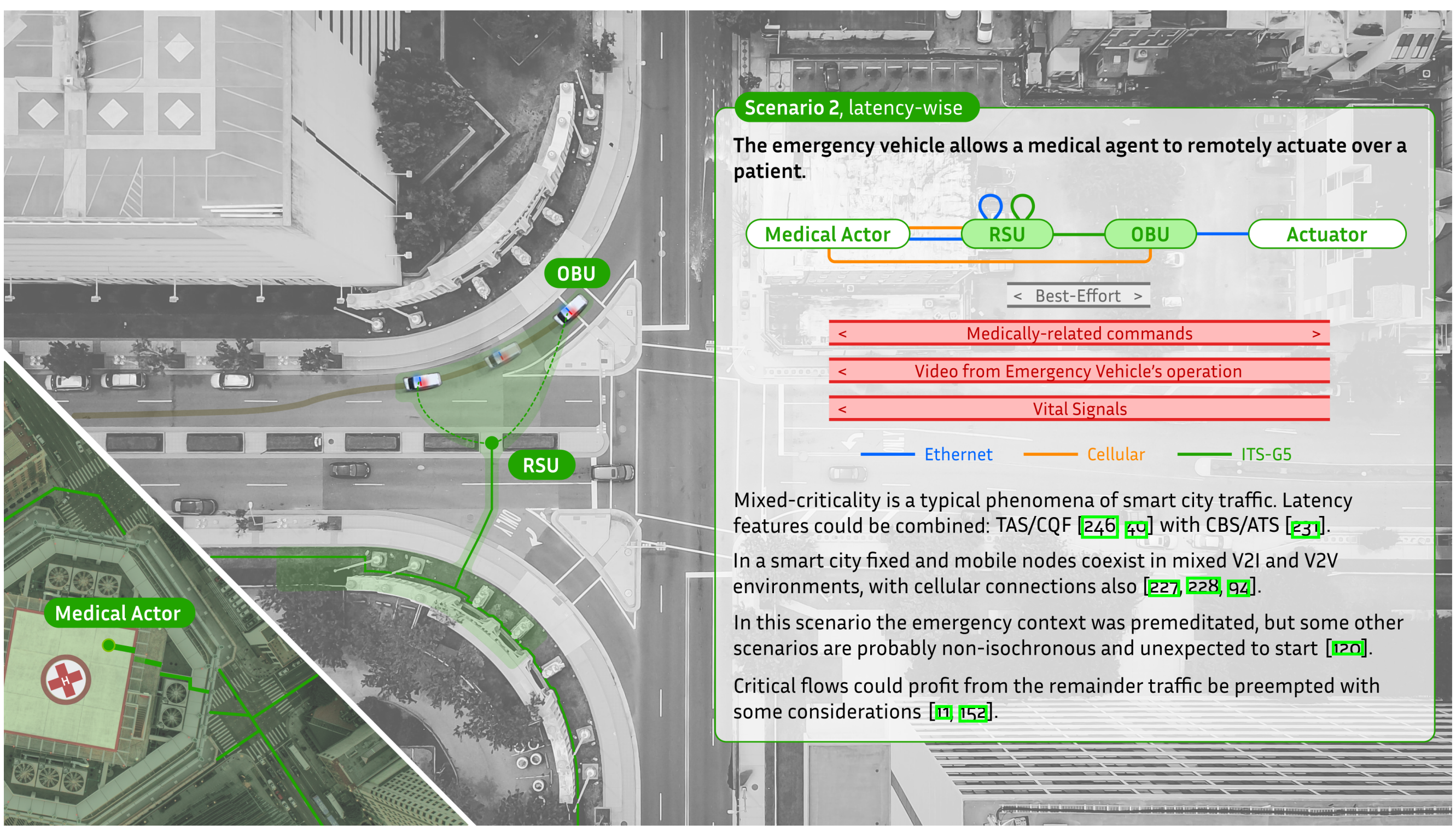

**FIGURE 11** Telehealth and remote features from a hospital to an ambulance use case scenario, in a latency perspective

## 5.6 | Discussion

From all the topics above in every standard and mechanism for latency in TSN, it is natural to understand why this pillar is so dense regarding the number of works found published by the scientific community during the years: the main criteria for choosing to deploy TSN paradigms are specified in the latency standards/mechanisms of TAS, CBS, ATS, FP, and CQF.

What has been specified in these items is not to be considered as alternatives between them but as complements to reach guarantees in terms of time, bandwidth, or other levels of criticality employing preemption. The scientific community has used all these standards/mechanisms as a key standard for achieving proper solutions to problems regarding strict guarantees and requirements in network transmissions. Even so, not all the listed standards are commonly used by the community, focusing more on using TAS (when there are strict time requirements to achieve), CBS (when there are strict bandwidth requirements to follow), and FP (when there are critical flows). There are also efforts to combine such shapers simultaneously [163, 219, 248]. The remaining standards are being recently considered for use in 5G-and-beyond cases.

This line of thinking allows us to transition into the smart city use case with insights into leveraging TSN to address the mixed-criticality typical of a city's diverse flow structure. To illustrate this concept, we turn to our motivation's second use case scenario (depicted in figure 11). Here, we observe three distinct types of high-priority traffic, each vying for time and bandwidth against other flows. It is imperative to characterize each traffic flow based on its specific requirements accurately—periodicity (if periodic), utilized bandwidth, or time constraints [140].

Unfortunately, if we look at what is implemented in the context of the Linux operative system, only some standards already have a ready-to-run version. While TAS and CBS have different implementations through the traffic control software (`tc`) as part of the more extensive `iproute2` package, through a queueing discipline (*qdisc*), ATS, CQF, and FP do not have a proper and straightforward implementation. Nonetheless, some vendors have already deployed some of these standards and mechanisms onto their equipment, some over Linux, but still prefer not to disclose their version to the public community.

Considering the versions already developed and currently set public to the Linux community, their configuration methods, despite not being the most straightforward nor intuitive, are exhaustive through the use of the Linux `iproute2`'s traffic control `tc` command, they can be very precise relative to what one might want to configure, from a schedule base-time to the definition of a gate control list, passing by controls of priority mappings or idle and sending slopes in the case of CBS. All the configuration parameters one can foresee when studying the possibility of adopting one or more of these standards through Linux can be described via these *qdisc* implementations. Nonetheless, there are hardware requirements that might be more strict to satisfy depending on the level of time precision needed.

However, while this implementation method is relatively straightforward to follow when one wants to statically configure a network node instance, in the case of dynamically configuring a set of nodes in a quite similar manner to an software-defined network (SDN) controller to SDN switches (as it will be mentioned next in the configuration category of TSN), Linux `iproute2`'s package does not provide a `tc` library to programmatically access the queueing disciplines directives of listing, adding, or removing rules, limiting the ways of conducting such a task through the usage of an own-made software. This limits us to at least two options for implementation: i) to use sub-processes and call `tc` from instantiated sub-shells, having our software depend on a set of shell calls, which is not a convenient way of solving such an issue, although it is quite common; ii) to implement such three directives via the usage of the under-the-hood `libnetlink` library, which can interact with the Linux kernel to handle and manipulate *qdiscs* and other network interface tasks.

As identified in this state-of-the-art latency study, most works attempt to calculate the correct schedules to apply to a given list of traffic flow characteristics in offline or online approaches. Mainly in the latter, the scientific community proves it difficult to reach a simple or more straightforward solution but fails to offer optimal solutions. As seen in table 6, for instance, there are already some efforts to apply heuristics that ease this process, but there is still plenty of room for the design of new abstraction layers that could bring these configurations to a more service-oriented level.

## 6 | STATE-OF-THE-ART IN CONFIGURATION COMPONENTS

Regarding TSN pillars, until now, we have brought standards and mechanisms that primarily work under a network made by its bridges. The approach now configures such a network and connects between bridges and end devices (such as talkers and listeners).

This reasoning gives body to a new and last pillar, which is the configuration, where the toolbox of TSN has standards such as of the stream reservation protocol (SRP) and the still-in-draft resource allocation protocol (RAP). Both standards have been actively debated and subject to new advancements in the scientific community.

### 6.1 | Stream Reservation Protocol (SRP)

Park et al., in [175], developed an automatic deregistration, employing a timer, for the SRP, which already features stream registration, reservation, and base deregistration functionalities for the transmission paths between a talker and a listener. SRP was also enhanced with three different TSN topologies: a fully centralized topology, a centralized network but distributed user model, and a fully distributed topology. In the latter, end-stations containing user streams communicate such user requirements directly over the TSN user/network protocol. Alternatively, a new component is added in a centralized network/distributed user topology model—the centralized network configuration (CNC),—an entity responsible for exchanging the configuration between talkers and listeners, offering a complete view of the network's physical topology. Lastly, with the fully centralized topology, another component is added—the centralized user configuration (CUC)—which allows users to be configured centrally, sharing data with the CNC. These topologies were analyzed in some works [211, 35, 163, 37, 203, 4, 39, 46, 236].

#### Topology models and Configuration Paradigms

Despite all the alternatives, choosing a distributed model is often less agile to changes in network configuration than centralized approaches, the reason it took several researchers to develop studies on how time-sensitive network topologies should be displaced [70, 68, 222, 69, 161, 163, 46, 156, 77, 134, 57]. Some of the found analyses were performed by using a logic programming framework (such as the works by Farzaneh et al. [70, 68, 69]), while others (such as of Thiele et al. [222]) use compositional performance analysis.

Apart from the modeling of time-sensitive network topologies, the TSN configuration method is also the subject of study, in which both user and network data must be exchanged between entities such as the CUC and CNC in the case of centralized topologies [95]. These methods are critical mainly for reconfiguration purposes in runtime.

In 2017 and 2018, Raagaard et al. [189] as well as Pop et al. [182], respectively, have performed studies on the runtime reconfiguration of TSN schedules through a heuristic algorithm such that the deadlines are satisfied. Queue usage is minimized, and non-critical traffic is accommodated.

In the standard where TSN's SRP is defined (IEEE 802.1Qcc [112]), an interface for configuration is made available to use of NETCONF and RESTCONF, whose interaction, to happen, requires data under the standardized format of YANG (acronym for yet another next generation, a data modeling language) defined in IEEE 802.1Qcp [113]. The latter authors already included this approach in their works, but several others also developed solutions with this approach, enabling a flexible network configuration model [91, 163, 203, 12, 236, 193, 167].

Between talkers/listeners and the centralized user configuration (CUC), in industrial contexts, it has been usual to find the presence of another protocol: the Open Platform Communications unified architecture (OPC-UA). OPC-UA is an industrial protocol and modeling standard that authors integrate with TSN to provide reliable network interoperability between communication layers of session, presentation, and application [206]. Several solutions were found studying this integration [182, 178, 35, 226, 203, 64, 135, 236, 57, 126, 167].

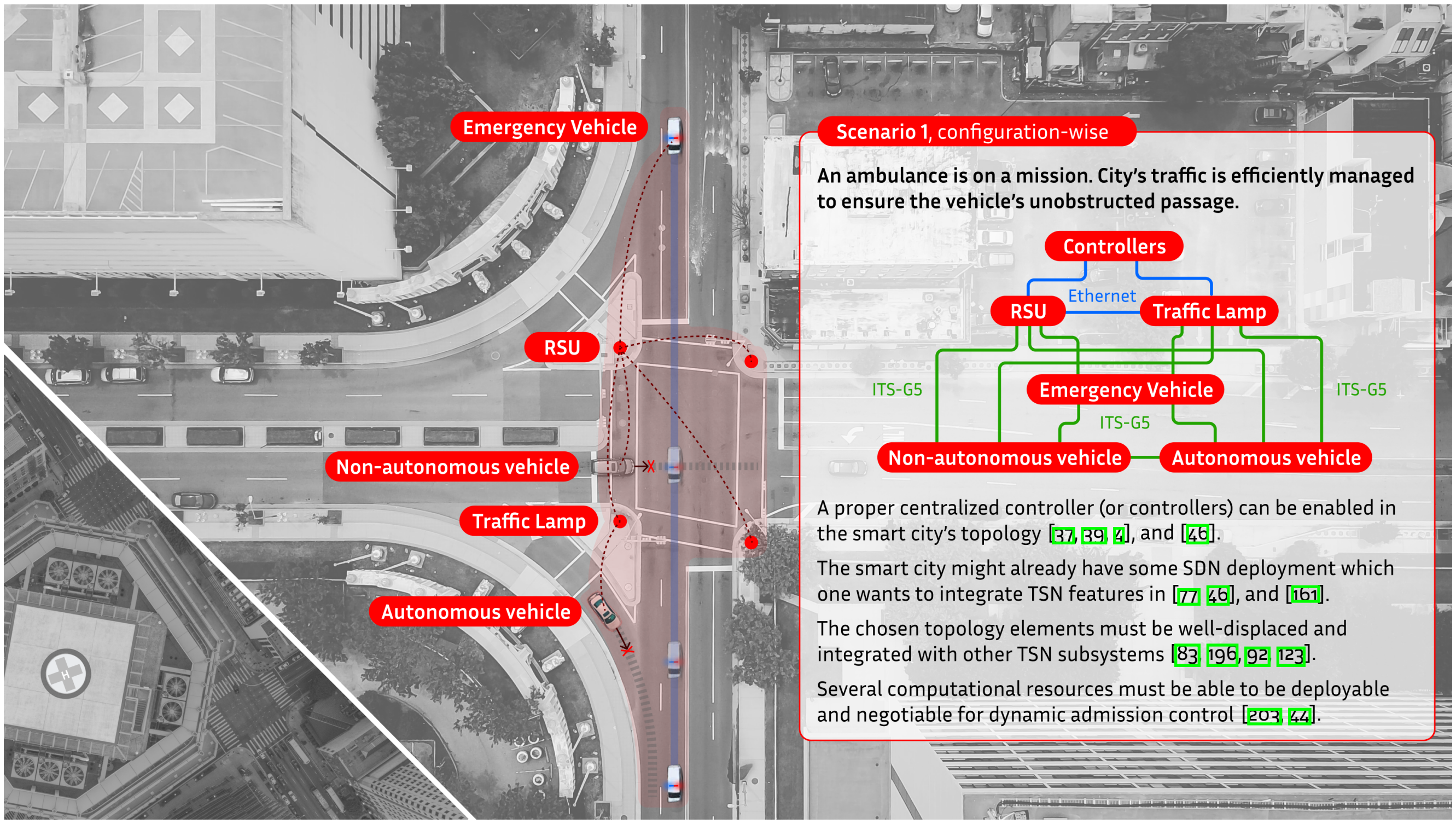

**FIGURE 12** Ambulance on a mission with reserved lane use case scenario, in a configuration perspective

Following a centralized-wise approach, and as such TSN deployments have a similar architecture to an SDN topology, several solutions have been found using SDN as a method for centralizing the configuration method and directives to all the network nodes in multiple TSN domains [166, 186, 37, 83, 92, 196, 155, 123, 39, 19, 44, 128, 236, 197, 36, 157, 125].

Summarizing all the related works to SRP, table 11 has a full-length list of articles sorted by categories.

**TABLE 11** Summary of all collected related work to SRP in TSN

| **1. Overall studies, context, and conceptualization** |
|---|
| [175], [163], [211], [35], [37], [203], [39], [4], [236], [95], [147], [177], [210], [5], [75], [167], [248] |
| **2. Models** |
| [70], [68], [222], [69], [163], [161], [77], [46], [134], [156], [57], [210], [75] |
| **3. Configuration methods** |

| i. Heuristics | iii. OPC-UA |
|---|---|
| [189], [182] | [182], [178], [35], [226], [135], [203], [64], [206], [236], [189], [126], [167] |
| **ii. YANG** | **iv. SDN** |
| [91], [163], [203], [236], [12], [193], [147], [167] | [166], [186], [37], [92], [83], [196], [39], [155], [123], [19], [128], [197], [44], [236], [36], [157], [20], [210], [125] |

## 6.2 | Resource Allocation Protocol (RAP)

Alternatively and compatible with the SRP, the resource allocation protocol (RAP) is still a draft standard. RAP, as detailed in the IEEE 802.1Qdd draft[†], is a hop-by-hop protocol for dynamic resource reservation based on the link-local registration protocol for transport purposes [172].

As it is a very recent advancement regarding the toolset of TSN, both studies and open implementations using RAP are less known of existence but identified as currently lacking the ability to reserve resources for time-aware scheduled streams in a centralized network/distributed user TSN topology model. Osswald et al., in [172], have described an extension to RAP with a first implementation published in the open-source community, which can be used as a basis for developing CNC connectors for specific use cases.

† https://1.ieee802.org/tsn/802-1qdd/

**TABLE 12** Summary of all collected related work to RAP in TSN

| 1. Overall studies, context, and conceptualization | 2. Extensions |
|---|---|
| [172], [203], [177], [210], [248] | [203], [89] |

In attempting to bridge time-sensitive networks with non-timely sensitive network devices, Grigorjew et al. [89] developed a minimalist reservation protocol addressing information on layers 3 and 4 to support the standardization efforts as of RAP. Other authors followed with similar contributions in RAP developments [203, 44].

Finally, and closing with RAP, in table 12, one can find a list of all the articles found related to such a standard.

## 6.3 | Discussion

From all the current state-of-the-art described before in terms of configuration, one can verify that multiple methods of configuring TSN deployments exist. Nonetheless, these configuration methods strongly depend on the adopted topology of what is being deployed.

In the case of a fully distributed topology for TSN, configuration methods do not vary much from a manual or static configuration, with the possibility of using YANG to share the parameters for configuration through NETCONF or RESTCONF methods. In the alternative, centralized topologies allow one or two components to take control of the network nodes (if one, it is related to network control—CNC; if two, then we are mentioning a component for network control and another for user configuration—CUC). As this shares a similar fashion to SDN plane separation between data and control, as mentioned, works are exploring using SDN to configure TSN deployments.

In fact, regarding the openness of TSN implementations per category of its standards, work in the configuration pillar of TSN is still quite far from the ideal stage. For instance, considering centralized topology models where the need for a centralized unit of CNC or CUC exists, there is currently no open-source implementation of a configurable CNC or CUC that is known to us in the public community. This might be an issue for those who intend to get as close as possible to a plug-and-play solution using the most open-source software possible, but although we are tending to reach such a spot, we are not there yet. Some vendors offer their implementations of a CNC (or even of a CUC), as in TTTech's[†] and Relyum's[‡], but they are not open regarding its implementation and use.

Stepping again into our smart city use case, configuring the TSN features will likely be one of the more intricate tasks. This complexity arises from the unique dynamics of a city's infrastructure, distinguishing it from conventional TSN use cases in the scientific and industrial realms. While we can characterize traffic within a city, the infrastructure accommodates many traffic flows in constant motion, necessitating dynamic path setups for their transmissions. Configuring time schedules, latency features, and reliability in such a dynamic scenario becomes challenging. For instance, within the context of the first use case scenario from our motivation (illustrated in figure 12), ensuring effective configuration strategies for a mobile node like an ambulance is crucial. These strategies must accommodate the node's movement, potentially utilizing detailed maneuver information transmitted and processed through the RSUs. Performing this will require that automatic or online methods for network traffic classification are created and used so that traffic can be correctly handled within the infrastructure nodes.

Regarding the configuration methods, as there still is no typical implementation to follow, the scientific community is not focused on a particular method from heuristics, methods using YANG, OPC-UA, or SDN. Not that these are meant to be alternative methods from one another, but there is still a road to pave to give both industry and the scientific community space to focus on a standardized fashion of delegating configuration rules and tasks to TSN nodes by using OPC-UA or YANG.

From these arguments, it is clear that we still need some advancements to state the configuration of a TSN network (independently of its topology) as an easy-to-follow and easy-to-maintain sequence of procedures. Moreover, the TSN mechanisms still need some openness to dynamic setups, such as dynamically set schedules in TAS or other rules in other shapers and filtering components. Answering this requirement, the TSN workgroup has already advanced onto an evolution of the SRP, the RAP, as mentioned earlier. While the SRP already has some implementations made, to the best of our knowledge, RAP still has not at least been made public.

# 7 | CONCLUSIONS AND FUTURE WORK

There is still a long way to pave in what time-sensitive networks can and will offer in the future, but we already have many features that can provide time-strict requirements to an Ethernet-based network. As mentioned earlier in this article, working in TSN is cautious in knowing when to transmit a specific flow rather than attempting to do our best to place a transmission and diminish latency and jitter values.

[†] https://www.tttech-industrial.com/products/slate/slate-xns/
[‡] https://www.relyum.com/web/rely-tsn-configurator-2/

## 7.1 | Discussion on Conclusions

TSN is a standardized set of tools and protocols that attempt to bring features of time and robustness to more common networking scenarios in how we forward content from node to node. This is a scenario that both industry and automotive have already been used to for many years since they have a clear view of all the aspects regarding their infrastructure and uses of such an environment. On the other hand, when we try to bring these concepts into a more proximate reality of a typical user, traffic and the number of connected devices change dynamically so that time-strict criteria are not adequately conditioned to work robustly.

One use case in which it is more common to have such types of challenges (regarding the dynamic changes in infrastructure and multitude of traffic flows) is the network infrastructure and usage within the concept of a smart city, to which this article was targeted to study if TSN conforms to solve its issues. Our main conclusion with our study is positive since there are plenty of examples of traffic flows within the use case of a smart city which could profit from having time-strict criteria but do have its issues, mainly in terms of configuration, regarding its highly dynamic topology and flow changes, which are currently not considered in TSN, looking over the common use cases of these approaches. Nonetheless, the TSN approach within this context would allow proactive solutions that would permit critical traffic to be transmitted without causing any trouble to the main functions to which the devices in the city were deployed to solve.

Adopting TSN in a scenario such as a smart city is similar to considering deploying both time synchronization and scheduling procedures, where the latter strongly requires traffic to be pre-known to be successfully classified in their requirements. As a smart city's infrastructure is not as closed as an industrial or automotive topology, difficulties arise in performing such an action since a large variety of factors influence this assessment. From the immense multitude of functions provided in the network to the highly dynamic topology changes to define a smart city set of traffic flows, these are complex tasks compared to the common set of flows of a TSN usual deployment.

Even considering a traffic flow classification completed, when configuring a standard such as of TAS or CQF, in which time schedules are to be applied, the task to optimize a better usage of a scheduling cycle is the main issue tackled by the works described in section 5.1, in which some of the alternatives point out to try to diminish the search domains for solutions with the help of heuristics, ILP and SMT methods, or other techniques such as of network calculus, source routing or even artificial intelligence and machine learning. The easiness of implementing a practical solution on common equipment or operative systems, as seen by the majority of these frameworks, is not high, and a service-oriented abstraction layer for configuring rules is lacking.

Within the approach for a smart city, one possible design for it could profit from having a permanently available time slot already pre-allocated for emergency transmissions, which could be modified over time if it was, indeed, in use in a current event in need for more resources: for instance, adjustments in TAS/CQF could be made if more time for transmissions is required, or in CBS/ATS if more bandwidth is a requirement. Doing the scheduling assuming these emergency transmissions is a way of applicability for a heuristic to reduce the searching domain of an optimized solution, which could offer a more rapid fashion to design a set of TAS rules.

## 7.2 | Future Work

Our base argument is advocating for the use of TSN to address timing needs in the realm of a smart city's M2M services. This approach positioned the smart city as an encompassing framework incorporating established TSN profiles. The integration of these profiles played a pivotal role in contemplating the deployment of TSN to define the potential for profiling the smart city as a distinct TSN use case application.

### Global future directions by TSN component

From the collection of selected surveys as described and detailed in section 2, we have gathered all the future directions and open challenges and have compiled our updated version of them in this section. Figure 13 starts by showcasing our gathered future work regarding the four components of TSN (synchronization, reliability, latency, and configuration). This plot shows us that most researchers are currently pointing out the configuration features of TSN. When TSN (and the late AVB) started, the main components to be invested in were latency and synchronization—this happened since latency had the most standards that had come from AVB, and the newer standards and mechanisms, with TSN, would require clock synchronization. As we have seen through sections 3 to 6, these topics are already advanced and minimally structured in most of TSN profiles, and the scientific community is now strongly investing in both reliability and configuration components.

In figure 13, although not denoted yet, a new feature was added: development. As the evolution and trends in the development and integration of TSN have grown in past years, solutions to develop and further deploy integrated solutions into critical M2M network infrastructures are more commonly appearing. This leaves the scientific community to reason on what is required to enhance the development of new standards in TSN and integrations in other systems or with different technologies. In the next section, we will describe surveyed open challenges of TSN, starting with this recent topic of development, showcasing the adoption rate of TSN solutions within the scientific community.

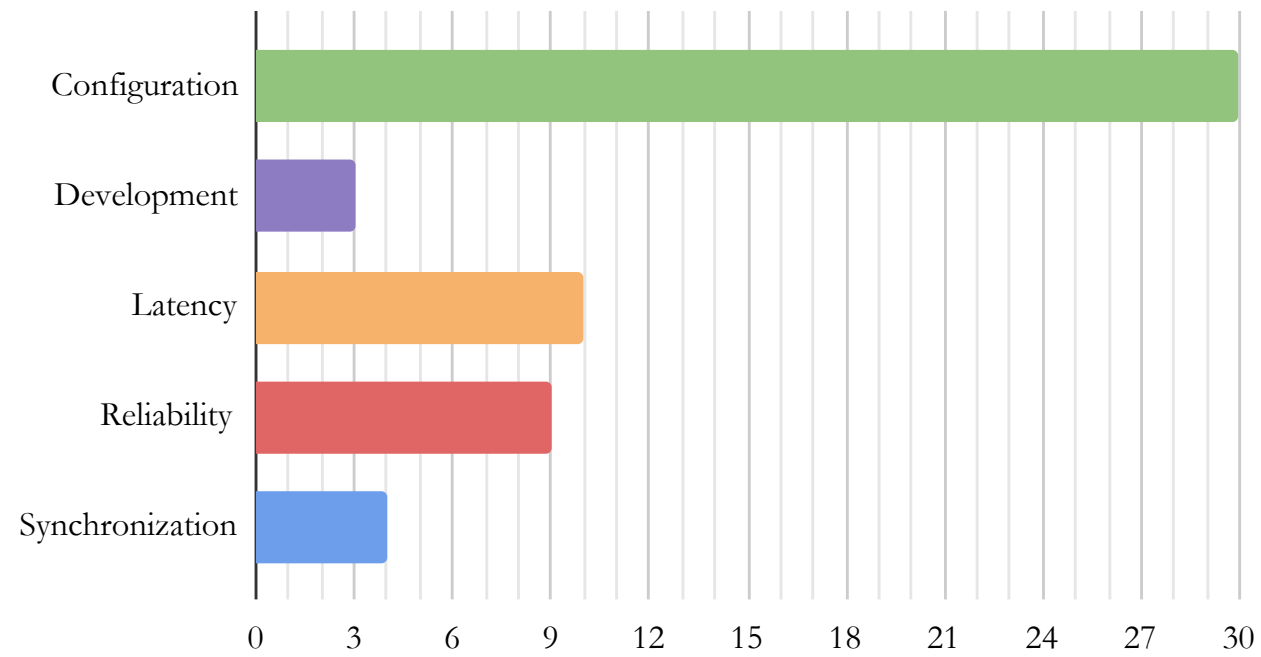

**FIGURE 13** Future work and open challenges per category, as collected in the selected state-of-the-art.

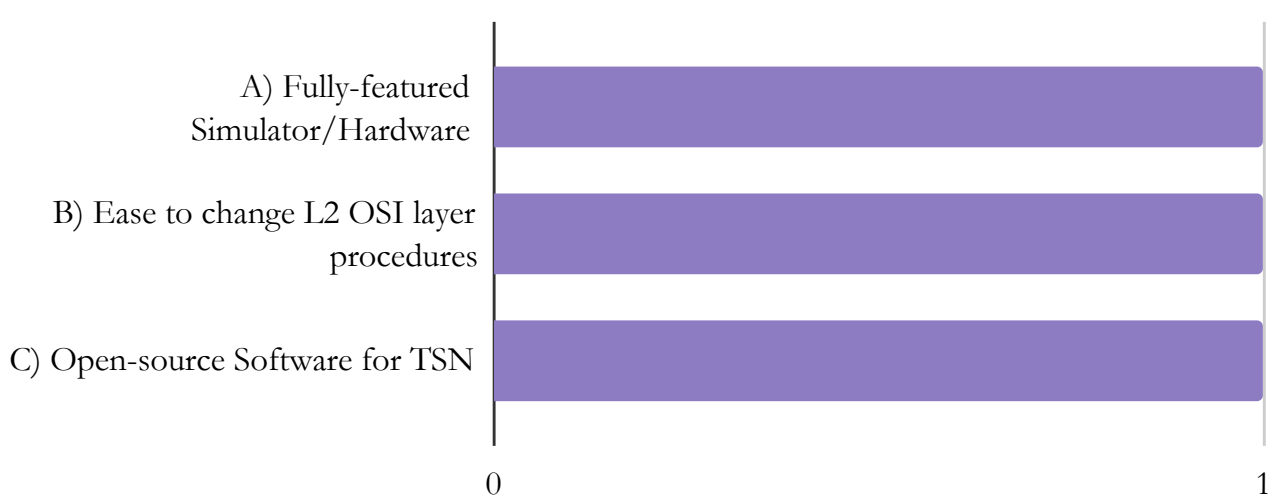

**FIGURE 14** Development-related future work and open challenges.

## Open challenges in Development

Regarding the development of solutions and integrations in TSN, as gathered from the selected surveys in table 1, we have collected three challenges as depicted in figure 14. Our vision of those challenges and our perspective of future directions are described below:

1) **Fully-featured Simulator/Hardware**
   Current TSN simulators lack comprehensive TSN features and realistic data models, hindering accurate evaluation. More advanced simulators and commercial off-the-shelf equipment are needed to advance TSN research and enable real-world deployments, such as the preparation for the smart city network environment. There are also no fully compliant hardware implementations enabling TSN to be experimented with in an open-sourced environment.
2) **Ease to change OSI Layer 2 procedures**
   The current methods to integrate or develop solutions in TSN require, sometimes, the introduction of complex mechanisms that change the OSI Layer 2 forwarding mechanism in a fundamental fashion, such as the case for implementing preemption or applying frame replication, which alters the protocol data units.
3) **Open-source Software for TSN**

**TABLE 13** Development open challenges survey references

| Challenge (as in fig. 14) | References |
|---|---|
| A) | [206] |
| B) | [90] |
| C) | [199] |

   Most solutions openly-available to TSN techniques are exclusively based on Linux's traffic control `tc` project. More open solutions are needed as alternatives to make TSN standards and mechanisms available to the public. Relatively to 5G, open-source software currently lacks support for TSN integration.
4) **Deployment of Solutions in Smart City Context**
   Within the context of a smart city, the network infrastructure is heterogeneous regarding computing node types and network access technologies. Moreover, the deployment varies not only per computing node type but also regarding scalability issues, whose difficulty will directly depend on the network infrastructure topology. A solution for distributing new versions or configurations along the infrastructure is an open problem.

Table 13 details the survey articles that mentioned each challenge in figure 14.

## Open challenges in Configuration

In terms of configuration, this TSN component is the most common category in the future directions as gathered in the selected surveys. The open challenges are described below:

1) **Automatic Traffic Classification**
   TSN achieves low latency and fault tolerance through traffic prioritization and shaping. However, real-time network traffic classification and status monitoring must adapt traffic class configuration without disrupting rules.
2) **Interoperability with Wireless Networks**
   In the scientific community, several solutions and studies on industrial TSN systems outline a roadmap for centralized configuration management of wireless TSN devices. More research is needed to address challenges like interference, unstable channel capacity, and interoperability between these environments, which lack a standardized procedure to execute it.
3) **Dynamic Reconfiguration**
   Dynamic reconfiguration, in centralized and fully distributed configuration management, lacks a standardized process specification, hindering interoperability. Improving real-time performance for fault recovery and reconfiguration requires further research in operating systems and network protocols. Moreover, to better handle the more variable traffic patterns

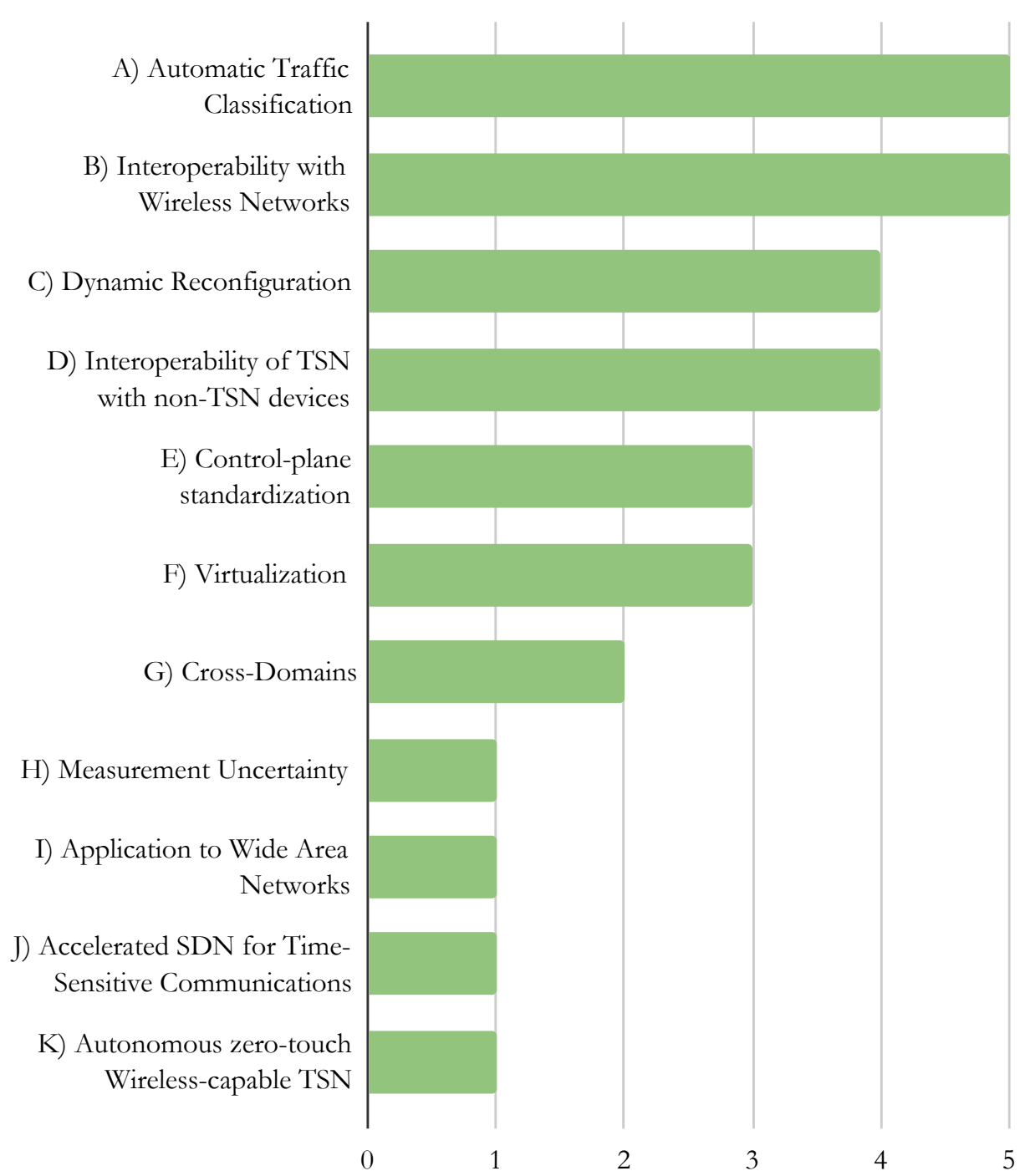

**FIGURE 15** Configuration-related future work and open challenges.

(such as smart cities), protocols or mechanisms should be designed to process admission requests for computational and network resources.

4) **Interoperability of TSN with non-TSN Devices**
Research on time-sensitive networks has focused on Ethernet links, neglecting interoperability with external networks, independently of its network access type. Future smart factories and vehicle networks require interoperability between heterogeneous technologies, including wireless communication systems like 5G, 6G, ZigBee, Bluetooth, Wi-Fi, dedicated short-range communication (DSRC), wireless access in vehicular environment (WAVE), and cellular vehicle-to-everything (C-V2X).

5) **Control-plane Standardization**
Standardization of TSN control plane needs improvement, particularly in defining CNC and CUC functions, unifying interface protocols, and enhancing the YANG model for unified configuration management.

6) **Virtualization**
As a factor for scalability and implementing different architectures and topologies, virtualization has been seen as a common deployment solution. There is a growing interest in integrating TSN with virtualization technologies like Kubernetes or considering wireless networks in architectures of NFV and MEC. This integration requires addressing interoperability, orchestration awareness of TSN assets, and exploring alternative traffic shaping algorithms and service function chains for dynamic environments. Wireless networks use hardware virtualization to provide customized QoS and flexibility. More research is needed on contact points between virtualized access networks and the TSN-based core network. Additionally, integrating TSN into the virtualized and interoperable O-RAN architecture for 5G RAN is a promising area for research.

7) **Cross-Domains**
Cross-domain TSN literature lacks specific technical details, particularly regarding east-west interface protocols and global controller design. Both approaches introduce challenges, including communication, negotiation, and clock compensation, especially in large-scale distributed networks.

8) **Measurement Uncertainty**
IEEE 802.11AS standard provides synchronization for Ethernet and Wi-Fi networks. While the synchronization activity is similar, media-dependent activities, such as timing message communication, differ between Wi-Fi and full-duplex Ethernet links. The IEEE 802.11 timing measurement (TM) procedure is used to calculate the propagation time, and the last version allows the use of a fine timing measurement (FTM) mechanism. This transposition on other TSN features in Wi-Fi is still an open research field [71].

9) **Application to Wide Area Networks**
Time-sensitive network protocols, effective in local environments, can also reduce end-to-end delay in wide area networks (WANs). However, managing TSN flows in large-scale WANs, especially for time-critical applications like telehealth, presents significant challenges.

10) **Accelerated SDN for Time-Sensitive Communications**
Research explores accelerating networking stacks for virtualized, time-sensitive applications, focusing on kernel bypass and userspace packet processing. While a data plane development kit (DPDK) is favored, an express data path (XDP) offers flexibility advantages, and further investigation is warranted. P4 programming language[†] shows promise for encapsulating TSN functionalities, warranting feasibility studies in clock synchronization, time synchronization message propagation, and network reconfiguration management.

11) **Autonomous Zero-touch Wireless-capable TSN**
Complementary proposals, including SDN-based network reconfiguration and machine learning algorithms, can optimize TSN and integrated 5G-TSN networks. Future research should focus on integrating wireless 5G domains, incorporating ultra-reliable low latency communications (URLLC) capabilities, and predicting varying network conditions for proactive management. These advancements can lead to substantial profits in 5G, 6G, and Industry 4.0 use cases.

[†] https://p4.org/

**TABLE 14** Configuration open challenges survey references

| Challenge (as in fig. 15) | References |
|---|---|
| A) | [163, 206, 219, 243, 248] |
| B) | [163, 210, 199, 243, 248] |
| C) | [56, 90, 210, 248] |
| D) | [163, 206, 56, 219] |
| E) | [163, 56, 210] |
| F) | [206, 199, 248] |
| G) | [42, 210] |
| H) | [71] |
| I) | [163] |
| J) | [199] |
| K) | [199] |

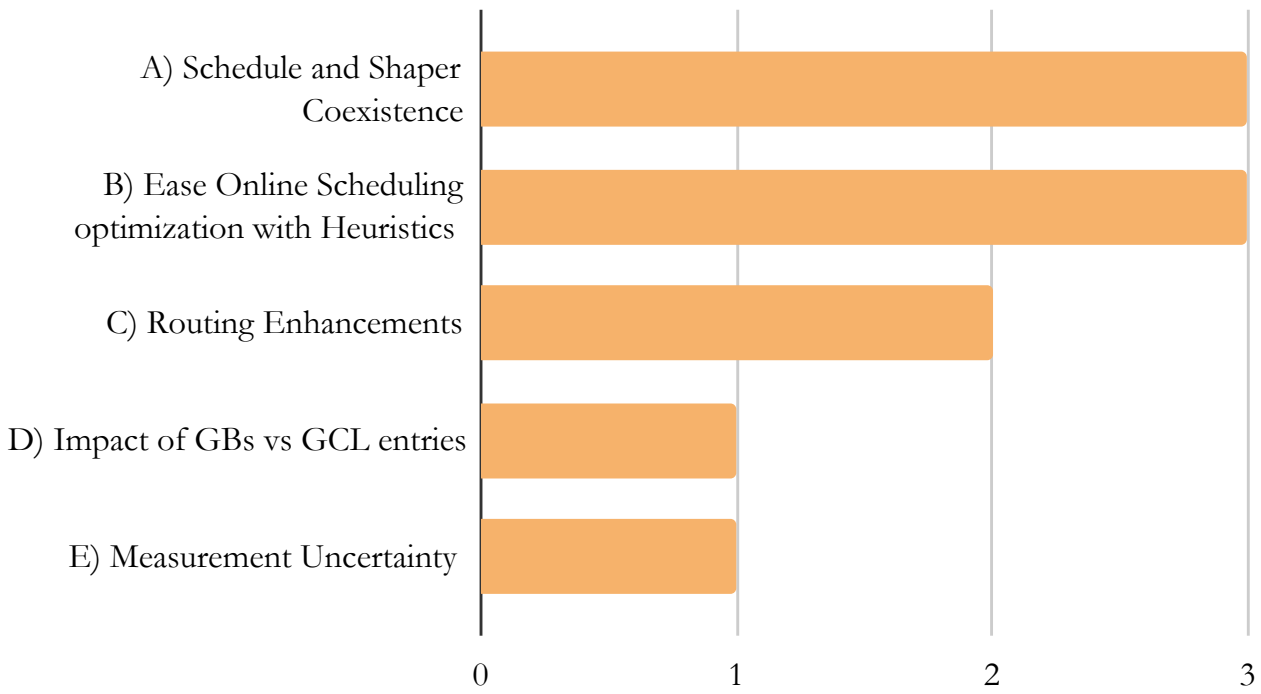

**FIGURE 16** Latency-related future work and open challenges.

Table 14 details the survey articles that mentioned each challenge in figure 15.

## Open challenges in Latency

In terms of latency, this TSN component is the most common category in the future directions as gathered in the selected surveys. The open challenges are described below:

1) **Schedule and Shaper Coexistence**
Current research on TSN shaping mechanisms primarily evaluates individual shapers, neglecting their combined use. Mostly in large-scale networks or contexts with many functions or services being provided, the combined use of shapers and schedules is still an open challenge.
2) **Ease Online Scheduling Optimization with Heuristics**
Online schedule reconfiguration, crucial for operations like automotive networks, is a computationally complex problem. While some preliminary works exist, more efficient reconfiguration algorithms are needed, especially for problem extensions.
3) **Routing Enhancements**
Joint routing and scheduling improve schedulability but are challenging to solve. While some algorithms handle multicast streams, research lacks insights into their schedule integration.
4) **Impact of Guard-Bands and Gate Control List Entries**
Literature lacks evaluation of guard-bands' impact on lower-priority traffic bandwidth. While AVB benefits from time-triggered stream scheduling with holes, this may require more gate closings and guard bands, potentially reducing available bandwidth. It's unclear how AVB and best-effort traffic can be integrated into a unified approach for scheduling time-triggered streams.
5) **Measurement Uncertainty**
Similarly to the case of measurement uncertainty in the configuration component open challenges, regarding latency features of TSN, there are still some considerations that are still open research problems when considering multiple network access technologies and dynamic traffic flows.

Table 15 details the survey articles that mentioned each challenge in figure 16.

**TABLE 15** Latency open challenges survey references

| Challenge (as in fig. 16) | References |
|---|---|
| A) | [163, 219, 248] |
| B) | [31, 42, 219] |
| C) | [206, 219] |
| D) | [219] |
| E) | [71] |

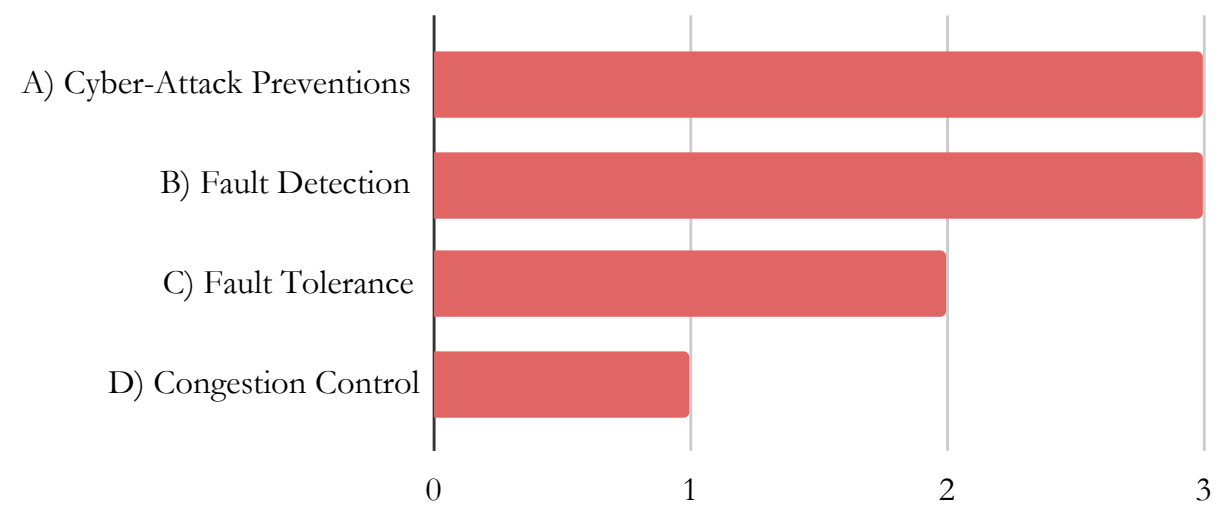

**FIGURE 17** Reliability-related future work and open challenges.

## Open challenges in Reliability

Regarding reliability, this TSN component is the most common category in the future directions, as gathered in the selected surveys. The open challenges are described below:

1) **Cyber-Attack Preventions**
Security in real-time Ethernet networks by means TSN has been overlooked, leaving vulnerable use cases like factory automation and in-vehicle networks at risk. Future standard protocols and mechanisms may incorporate security measures such as key exchange and source authentication. Some of these issues could already be filtered out with the help of standards such as PSFP, which can also enforce security prevention rules for jamming and denial-of-service attacks.

**TABLE 16** Reliability open challenges survey references

| Challenge (as in fig. 17) | References |
|---|---|
| A) | [56, 90, 219] |
| B) | [56, 219, 248] |
| C) | [163, 90] |
| D) | [56] |

2) **Fault Detection**
PSFP is not yet commonly implemented in commercial off-the-shelf equipment but could prevent schedule violations due to stronger control regarding unknown packet flows or other external interferences. However, its implementation may require a joint approach due to limitations on filtering entries in bridges.

3) **Fault Tolerance**
FRER enhances TSN frame transmission reliability by sending duplicate frames over disjoint paths. This primarily prevents packet loss, granting delivery under redundant paths with packet replication configurable by traffic class and path information. Currently, no publicly available sources exist for an implementation on FRER.

4) **Congestion Control**
Congestion control is crucial for achieving communication requirements and ensuring the efficacy of the programmed schedules without external interferences. While used and effective, deep learning models face challenges due to insufficient congested sample streams during training, impacting their congestion control capability.

Table 16 details the survey articles that mentioned each challenge in figure 17.

## Open challenges in Synchronization

In terms of synchronization, this TSN component is the most common category in the future directions, as gathered in the selected surveys. The open challenges are described below:

1) **Cross-Domains**
Clock mismatch in TSN networks, especially with multiple domains, poses a great challenge. Potential solutions include a synchronization controller at domain borders or time awareness in the control plane. Moreover, work is still being done on the proposed amendment of IEEE P802.1ASdm (Hot Standby), in which statically configures at least two gPTP grandmaster nodes for two independent time domains but tightly synchronized. Notwithstanding, cross-domain issues are also present when dealing with wired-to-wireless and wireless-to-wired transitions, where different and non-standardized means for maintaining time synchronization are due to cooperate.

2) **Clock Source Security Enhancement**
In the current form, as standardized, there is no formal way to ensure the identity of a gPTP grandmaster (or time transmitter) clock source. From this issue, cyber-attacks could originate, allowing external interferences to happen and, consequently, creating a snowball effect on other mechanisms (such as shapers) that strictly depend on the state of time synchronization.

3) **Tolerance to Byzantine Failures**
Current time synchronization protocols lack analysis of failure modes and quantitative reliability analysis. Improving reliability in the face of Byzantine failures and ensuring synchronization in large-scale networks are key areas for further study.

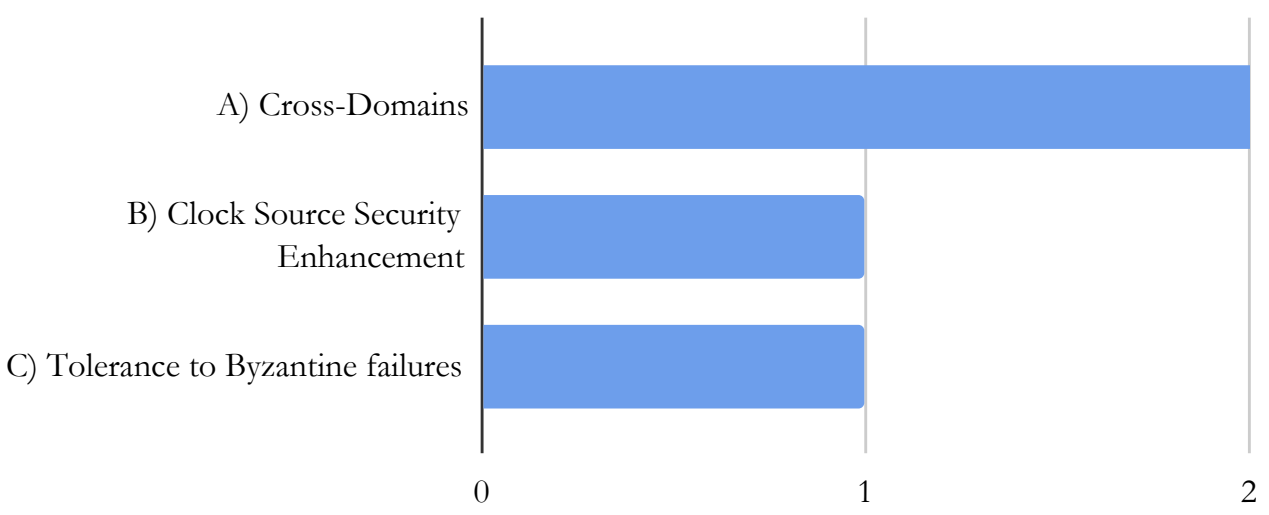

**FIGURE 18** Synchronization-related future work and open challenges.

**TABLE 17** Synchronization open challenges survey references

| Challenge (as in fig. 18) | References |
|---|---|
| A) | [42, 210] |
| B) | [90] |
| C) | [56] |

Table 17 details the survey articles that mentioned each challenge in figure 18.

## A future deployment in a smart city

Incorporating the concept of TSN within a smart city context is already demanding. However, the seamless communication between a diverse array of network elements, spanning both wired and wireless nodes, introduces another dimension that presents a challenging prospect for future development. As revealed in the literature review detailed from section 3, there have been efforts to explore the integration of TSN in wireless environments, including 5G and other technologies. The primary challenge in these endeavors lies in establishing proper signaling mechanisms to maintain time synchronization and configurations while facilitating redundant communication paths. This endeavor could potentially open the door to applying reliability features such as FRER in city infrastructures characterized by network topologies that lack cycles.

The intercommunication between different network access technology nodes could provide the network with redundant paths and extend the time requirements to network elements that Ethernet capabilities cannot characterize. For instance, while there are considerations of TSN within the scope of automotive use cases, the interaction between a vehicle (OBU) and an RSU cannot be currently performed with TSN considerations, since its connection must be made with wireless resources with non-Ethernet capabilities such as Wi-Fi or other technologies such as ITS-G5, or even 5G.

Vertically speaking, likewise, we can elevate ourselves overlooking to the smart city as an integration of several current TSN use cases still lacking some integration between one another, one can also be more proximate to the citizen itself in new applications. The smart city concept can then extend to applications that directly benefit citizens, such as smart hospitals (as hinted at in our second use case scenario in section 1.1). In these settings, TSN has the potential to offer solutions for future telehealth challenges[†].

Currently, transforming all nodes in a city into TSN-capable nodes (as of bridges or other end-nodes) could be a cumbersome task since both the hardware and deployment price and development cost for a solution are quite high. The hardware currently available on the market, and carrying the ability to run TSN procedures and standards, is still very proprietary and closed, which does not allow the scientific community and, consequently, other common users to profit from TSN-nodes quickly. Independently of that, there is already a large set of common devices carrying Intel NICs which are TSN-capable processing units[‡], but whose software usage does not take advantage of such properties. One example of this issue is the Aveiro Tech City Living Lab [192][§], in which we already have a large set of PcEngine's APUs deployed throughout the city of Aveiro, Portugal, which have (or at least most of them have) Intel's I210 NICs that are not currently used by any TSN procedure or standard. Our future work goal is to use this infrastructure to study our first TSN deployment in the context of a smart city in coordination with our SDN deployment, whose challenges were described earlier in this document.

Moreover, looking at a smart city infrastructure already deployed, it is relevant to consider the cross-compatibility between TSN-capable and non-capable devices. In such a scenario, one can then append specific TSN nodes, such as switches, with open ports to configuration procedures, as described in section 6. With such a capability, network nodes and all of the TSN features could be configured regarding a chosen architecture, which leads to the possibility of having multiple controller units (CNCs, or CNCs and CUCs).

The architectural choice for a TSN deployment within a smart city could be a strategic decision regarding critical services the infrastructure is supposed to provide. As there is no standardized topology for a smart city network, the TSN architecture must be chosen according to the main requirements of the traffic to be timely scheduled or subject to latency and jitter levels.

For instance, as soon one mentions an emergency as one of the services to be provided, it makes sense for the network to be capable of providing redundant resources to enhance its availability levels. In some topologies, this could be a difficult target to reach since the infrastructure might not allow cycles or redundant paths to form. Nonetheless, considering different and non-Ethernet derivative communications alternatives, one could profit from multiple paths to the same target. Following this reasoning, and since this is a current research topic in TSN, there is the need to attempt to integrate and extend TSN domains to outside Ethernet. This is another future work for anyone applying a TSN approach to a network that coexists with wireless technologies.

Within the context of a smart city, this approach of introducing TSN features into wireless environments could also allow vehicles carrying OBUs to participate in the network, providing means for an extension of the city TSN domain into an inter-network of vehicles, allowing, for example, for intra-vehicle messages to trigger events on their current environments. Notwithstanding, cellular connections such as 5G could be integrated and interfaced with TSN, as there are already some developments regarding this topic, to allow redundant or alternative connections and enhance ways to implement reliability into an infrastructure. There are already some efforts regarding implementing TSN-compatible solutions in integration with both 5G or WiFi 7, as shown throughout this paper, but no formal equivalence or standardization exists.

On a larger scale, one can also foresee inter-city communication, where messages are communicated in-between cities and no TSN or other qualified network equipment exists. To allow this scenario, TSN-compliant network infrastructures must exist in each city, and the cross-domain TSN open issues, as described earlier, are required to have a solution.

[†] Intel has already started working in this type of application. For more information, consult https://www.intel.com/content/www/us/en/healthcare-it/smart-hospital.html (accessed in September 2024).

[‡] https://www.intel.com/content/www/us/en/developer/articles/technical/adopting-time-sensitive-networking-tsn-for-automation-systems-0.html (accessed in September 2024).

[§] Online platform available in https://aveiro-living-lab.it.pt.